\documentclass{iopjournal}

\usepackage{amsmath}
\usepackage{amssymb}
\usepackage{amsfonts}
\usepackage{bm}
\usepackage{mathtools}
\usepackage{braket}
\usepackage{graphicx}
\usepackage{hyperref}
\usepackage{amsthm}

\newtheorem{theorem}{Theorem}
\newtheorem{corollary}{Corollary}

\newcommand{\Tr}{\operatorname{Tr}}
\newcommand{\Ree}{\operatorname{Re}}
\newcommand{\Imm}{\operatorname{Im}}
\newcommand{\dd}{\mathrm{d}}
\newcommand{\ii}{\mathrm{i}}
\newcommand{\id}{\mathbb{I}}
\newcommand{\calL}{\mathcal{L}}
\newcommand{\calF}{\mathcal{F}}
\newcommand{\calA}{\mathcal{A}}
\newcommand{\calB}{\mathcal{B}}
\newcommand{\calJ}{\mathcal{J}}
\newcommand{\calD}{\mathcal{D}}

\newcommand{\calE}{\mathcal{E}}
\newcommand{\calH}{\mathcal{H}}

\newcommand{\calV}{\mathcal{V}}
\newcommand{\calQ}{\mathcal{Q}}
\newcommand{\out}{\mathrm{out}}
\newcommand{\inn}{\mathrm{in}}
\newcommand{\ssr}{\mathrm{ss}}
\newcommand{\T}{\mathrm{T}}

\begin{document}
\articletype{Paper} 

\title{Finite-Frequency Fluctuation-Response Bounds for Open Quantum Systems}

\author{Jie Gu$^1$, Kangqiao Liu$^2$}

\affil{$^1$Chengdu Academy of Educational Sciences, Chengdu 610036, China}

\affil{$^2$School of Science, Key Laboratory of High Performance Scientific Computation, Xihua University, Chengdu 610039, China}

\email{jiegu1989@gmail.com}


\begin{abstract}
We derive a finite-frequency fluctuation-response inequality for Markovian open quantum systems in an input-output setting. For any downstream measurement of the emitted field, the measured lock-in response-to-noise matrix is bounded by the output-field quantum Fisher information rate. For dissipative amplitude modulation with vacuum inputs, this information rate is further bounded by a frequency-independent signal-channel activity, which reduces for kinetic modulation to the stationary channel fluxes. For a faithful stationary state, complete access to the Markov output, and horizontal dissipative tangents, we also evaluate the intermediate output-field quantum Fisher information rate explicitly and show that it equals the activity matrix. The result is detector-facing but unraveling-independent: it applies after choosing a measurement record, while the information ceiling is set by the quantum field before any detection scheme or trajectory representation is selected. We formulate the bound for multiple signal channels and real finite-frequency quadratures, and illustrate it with a single-sided cavity, resonance fluorescence, and a two-port truncated Kerr-parametric cat resonator exhibiting appreciable channel mixing and off-diagonal matrix effects.
\end{abstract}


\section{Introduction}

Weak periodic signals are commonly detected through the fields emitted by an open quantum device.  This is the case in quantum optics, circuit QED, optomechanics, resonance fluorescence, and mesoscopic transport, where the experimentally recorded object is typically an output current: a homodyne or heterodyne photocurrent, a photon-counting trace, or a filtered electrical current \cite{ClerkRMP,BlanterButtiker2000,WisemanMilburn}.  At a fixed analysis frequency, the central quantities are therefore not equal-time variances but spectra and lock-in response coefficients \cite{DechantLutz2025Sensing}.  A natural question is then how large a coherent response at frequency $\omega$ can be, relative to the spontaneous fluctuations seen in the same detector record and relative to the physical activity through which the signal is injected.

At thermal equilibrium, the fluctuation-dissipation theorem answers this type of question by relating linear response to equilibrium fluctuations \cite{CallenWelton1951,Kubo1957,Kubo1966}.  Away from equilibrium, the equality is generally lost, but response and fluctuations remain constrained by information-theoretic principles \cite{MarconiPuglisiRondoniVulpiani2008,BaiesiMaesWynants2009,SeifertSpeck2010}.  Classical fluctuation-response inequalities express this idea through the Cram\'er--Rao bound or relative-entropy bounds on stochastic path measures: a large response requires either large fluctuations in the measured observable or large distinguishability between the unperturbed and perturbed dynamics \cite{Rao1945,Cramer1946,DechantSasa2018,OwenGingrichHorowitz2020,FernandesMartinsHorowitz2023,AslyamovEsposito2024,GaoChunHorowitz2024,PtaszynskiAslyamovEsposito2024}.  Finite-frequency versions of this principle have recently been developed for classical Markovian dynamics, including jump processes and Langevin systems \cite{Dechant2025,ZhengLu2026}.  Related spectral bounds constrain deviations from equilibrium-style fluctuation-dissipation predictions in nonequilibrium Markov processes \cite{HaradaSasa2005,HaradaSasa2006,DechantPSD2023,GuFDRI2026}.  These results provide a well-defined classical framework for response precision at finite frequency.

Open quantum systems add a further layer that is essential for detector-facing bounds.  Before any classical trajectory record is obtained, the system has emitted a quantum output field.  Different detection schemes correspond to different positive-operator-valued measurements on this field.  Photon counting, homodyne detection, heterodyne detection, and more general filtered measurements are therefore not merely different coarse grainings of a single classical trajectory; they are different downstream measurements of the same emitted quantum field \cite{Carmichael,Dalibard1992,BoutenHandelJames2007}.  A bound applied after choosing a particular quantum trajectory is operational for that detector, but it is not independent of the unraveling \cite{VanVu2025QuantumTrajectory}.  Conversely, a bound written only in terms of internal noncommutative correlation functions need not correspond directly to the spectra and response coefficients displayed by a laboratory receiver.

This paper formulates a finite-frequency fluctuation-response bound at the quantum input-output level.  The emitted field is treated as the measurement-independent carrier of information, while the observed detector current is a classical record obtained from that field by a chosen measurement and subsequent filtering.  The result is therefore detector-facing but unraveling-independent: its left-hand side contains only measurable output-current spectra and lock-in responses, while the upper limit is imposed before selecting photon counting, homodyne detection, heterodyne detection, or any other downstream measurement.

The central statement, which does not require a matrix-valued long-time limit of the output-field quantum Fisher information, is the directional inequality
\begin{equation}
\vartheta^{\T}
\mathsf R^{\T}(\omega)
\left[\mathsf S^{\out}(\omega)\right]^{+}
\mathsf R(\omega)
\vartheta
\leq
\overline{\calF}^{Q}_{\out}(\omega;\vartheta)
\leq
\vartheta^{\T}
\left(\calA_{\rm sig}\otimes\id_2\right)
\vartheta
\label{eq:introbound}
\end{equation}
for every real signal direction \(\vartheta\).  Here \(\mathsf S^{\out}(\omega)\) is the covariance matrix of the measured real lock-in current modes, \(\mathsf R(\omega)\) is the corresponding finite-frequency response matrix, and \(\mathsf{M}^+\) denotes the Moore--Penrose inverse of a matrix \(\mathsf{M}\).  The intermediate directional quantity \(\overline{\calF}^{Q}_{\out}(\omega;\vartheta)\) is the upper long-time quantum Fisher information rate carried by the emitted output field along \(\vartheta\).  The matrix \(\calA_{\rm sig}\) is the signal-channel activity associated with calibrated dissipative amplitude modulation of the Markovian coupling operators, and the factor \(\id_2\) acts on the cosine and sine components of the sinusoidal perturbation.

If, in addition, the ordinary matrix limit
\begin{equation}
\calF_{\out}^{Q}(\omega)
=
\lim_{T\to\infty}T^{-1}F_{\out}^{Q,T}(\omega)
\label{eq:intro_qfi_limit}
\end{equation}
exists, then Eq.~\eqref{eq:introbound} is represented by the Loewner matrix chain
\begin{equation}
\mathsf R^{\T}(\omega)
\left[\mathsf S^{\out}(\omega)\right]^{+}
\mathsf R(\omega)
\preceq
\calF^{Q}_{\out}(\omega)
\preceq
\calA_{\rm sig}\otimes\id_2 .
\label{eq:introbound_matrix}
\end{equation}
Thus the unconditional result is Eq.~\eqref{eq:introbound}; the matrix form \eqref{eq:introbound_matrix} is used only when the limit \eqref{eq:intro_qfi_limit} exists.  These inequalities state that no output detector can display a response-to-noise larger than the information available in the emitted field, and that this field information is itself limited by the activity of the channels through which the signal enters.

The formulation is intended to match how finite-frequency sensing is performed in quantum technologies.  The response matrix and the noise spectrum can be extracted from the same detector record using the same lock-in convention.  The activity matrix is fixed independently by the calibrated signal-bearing coupling channels.  Thus Eq.~\eqref{eq:introbound} gives a consistency relation between measured spectral signal-to-noise and the physical resources used to encode the signal.  It applies to arbitrary downstream measurements of the output field and to  any fixed-frequency linear lock-in filter, provided the open-system dynamics is Markovian, the perturbation is weak and sinusoidal, and the relevant signal channels are properly accounted for.

The result also clarifies the relation to response kinetic uncertainty relations.  Response KURs and their quantum open-system extensions bound response precision by a Fisher-information/activity mechanism \cite{LiuGuRKUR,LiuGuQRKUR}.  The present work places that mechanism at the level of the emitted field rather than at the level of a chosen measurement record.  This change is crucial in quantum input-output settings, because different detectors access incompatible aspects of the same output field.  The bound therefore gives a common information ceiling for homodyne, heterodyne, photon-counting, and more general detection architectures.

Beyond the general inequality, the paper develops a multiparameter matrix formulation for several modulated signal channels and for the two real quadratures of a periodic drive.  It gives explicit detector-level response-to-noise bounds for output currents, connects the activity term to the calibrated dissipative coupling tangents, and evaluates the intermediate full-output QFI for horizontal kinetic tangents.  
The examples include a Gaussian coherent-input benchmark based on a single-sided cavity, a finite-dimensional resonance-fluorescence example comparing detector Fisher information directly with the output-field QFI, and a two-port truncated Kerr-parametric resonator displaying substantial responses in both signal channels and nontrivial off-diagonal matrix blocks.

The paper is organized as follows.  Section~\ref{sec:setup} introduces the Markovian input-output model and states the main finite-frequency bounds.  Section~\ref{sec:remarks} discusses their operational interpretation.  Section~\ref{sec:examples} presents the three examples.  Section~\ref{sec:connection} relates the result to classical finite-frequency fluctuation-response inequalities, spectral fluctuation-dissipation-response bounds, response KURs, quantum Cram\'er--Rao theory, input-output theory, and quantum trajectories.  Section~\ref{sec:discussion} summarizes the limitations and possible extensions.  The detailed derivation is given in the Appendix.

\section{Main results}
\label{sec:setup}

We consider a finite-dimensional quantum system coupled to Markovian bosonic input fields through channels indexed by $\mu=1,\ldots,n_c$. Writing
\begin{equation}
\calD[L]\rho
=
L\rho L^\dagger
-
\frac{1}{2}\{L^\dagger L,\rho\},
\end{equation}
the unperturbed reduced state obeys the Gorini-Kossakowski-Sudarshan-Lindblad equation \cite{Gorini1976,Lindblad1976}
\begin{equation}
\dot\rho=\calL\rho
=
-\ii[H,\rho]
+
\sum_{\mu=1}^{n_c}\calD[L_\mu]\rho.
\label{eq:lindblad}
\end{equation}
We assume throughout the main theorems that the finite-dimensional
Markov semigroup is exponentially mixing.  More precisely, let
\[
\mathfrak T_0=\{X:\Tr X=0\}
\]
be the traceless subspace, and assume that there is a constant
\(\gamma_{\rm mix}>0\) such that
\begin{equation}
\operatorname{spec}\!\left(\calL|_{\mathfrak T_0}\right)
\subset
\{z\in\mathbb C:\Ree z\leq -\gamma_{\rm mix}\}.
\label{eq:mixingassumption}
\end{equation}
This assumption implies uniqueness of the stationary state
\(\rho_{\ssr}\) and exponential convergence to it.  It is slightly
stronger than mere uniqueness of a stationary state, but it is the
natural condition under which the stationary output spectra, finite-time
lock-in limits, and Liouvillian resolvents used below are unambiguous.
In particular, for every real \(\omega\) the operator
\(
-\ii\omega-\calL
\)
is invertible on \(\mathfrak T_0\).  All unperturbed spectra and
responses are evaluated in the stationary state \(\rho_{\ssr}\).
With \(L_\mu(t)\) denoting the Heisenberg-picture coupling operator, the
input and output fields satisfy \cite{GardinerCollett1985,GardinerZoller}
\begin{equation}
b_{\mu,\out}(t)=b_{\mu,\inn}(t)+L_\mu(t).
\label{eq:inputoutput}
\end{equation}
Unless stated otherwise, the incoming fields are in vacuum.

A downstream detector performs a measurement on the output field and returns real currents
\begin{equation}
\mathbf I(t)=(I_1(t),\ldots,I_m(t))^\T.
\end{equation}
This notation includes homodyne, heterodyne, counting, inefficient, and adaptive measurements. The resulting classical process need not be Markovian.

The signal is a weak time-dependent modulation of the coupling amplitudes. For $n_p$ signal parameters, write $\epsilon(t)=(\epsilon_1(t),\ldots,\epsilon_{n_p}(t))$ and
\begin{equation}
L_\mu^{(\epsilon)}(t)
=
L_\mu+
\sum_{q=1}^{n_p}\epsilon_q(t)M_{\mu q}
+
O(\|\epsilon\|^2),
\label{eq:generalcouplingmod}
\end{equation} 
where \(M_{\mu q}\) is the tangent of channel \(\mu\) with respect to
parameter \(q\).  The activity theorem below is stated for purely
dissipative amplitude tangents.  In the fixed input-output
representation used here, this means that the first-order tangent does
not contain an additional Hamiltonian-like Stinespring component:
\begin{equation}
\sum_\mu
\left(
L_\mu^\dagger M_{\mu q}
-
M_{\mu q}^\dagger L_\mu
\right)
=0,
\qquad q=1,\ldots,n_p .
\label{eq:purediss_tangent}
\end{equation}
If this condition is not imposed, a coherent channel tangent appears in
the sequential channel Fisher information.  Such a term is not bounded
by the jump activity alone and should be treated as a Hamiltonian signal
contribution.

The kinetic modulation used in several examples is
\begin{equation}
L_\mu^{(\epsilon)}(t)
=
\exp\left[\frac{1}{2}\sum_q b_{\mu q}\epsilon_q(t)\right]L_\mu,
\label{eq:kineticmod}
\end{equation}
with real calibration coefficients \(b_{\mu q}\).  In this case
\(M_{\mu q}=b_{\mu q}L_\mu/2\), and the purely dissipative condition
\eqref{eq:purediss_tangent} is automatically satisfied.

At a fixed nonzero frequency $\omega$, we use real unit-RMS envelopes
\begin{equation}
\phi_c(t)=\sqrt{2}\cos\omega t,
\qquad
\phi_s(t)=\sqrt{2}\sin\omega t,
\label{eq:unitrmsmodes}
\end{equation}
for which $T^{-1}\int_0^T\phi_\alpha(t)\phi_\beta(t)\dd t\to\delta_{\alpha\beta}$ with $\alpha,\beta\in\{c,s\}$. The local finite-frequency signal is
\begin{equation}
\epsilon_q(t)=\eta_{q,c}\phi_c(t)+\eta_{q,s}\phi_s(t),
\label{eq:realmodel}
\end{equation}
and the real parameter vector is ordered as
\begin{equation}
\eta=(\eta_{1,c},\eta_{1,s},\ldots,\eta_{n_p,c},\eta_{n_p,s})^\T.
\end{equation}
For a one-dimensional signal direction we write $\eta=\epsilon\vartheta$, where $\vartheta$ is a fixed real vector in this same space.

For each measured current component, define the finite-time positive-frequency mode
\begin{equation}
\tilde I_{a,T}(\omega)
=
\frac{1}{\sqrt{T}}
\int_0^T\dd t\,
e^{\ii\omega t}
\left[I_a(t)-\langle I_a\rangle_{\ssr}\right].
\label{eq:fouriercurrent}
\end{equation}
The real lock-in vector is
\begin{equation}
\mathsf X_T(\omega)
=
\left(
\sqrt2\Ree\tilde I_{1,T},\sqrt2\Imm\tilde I_{1,T},
\ldots,
\sqrt2\Ree\tilde I_{m,T},\sqrt2\Imm\tilde I_{m,T}
\right)^\T,
\end{equation}
with all Fourier modes evaluated at the same $\omega$. Its stationary noise matrix is
\begin{equation}
\mathsf S^{\out}(\omega)
=
\lim_{T\rightarrow\infty}\operatorname{Cov}\,\mathsf X_T(\omega).
\label{eq:realoutspectrum}
\end{equation}
The finite-frequency response matrix $\mathsf R(\omega)$ is defined by
\begin{equation}
\delta\langle \mathsf X_T(\omega)\rangle
=
\sqrt{T}\,\mathsf R(\omega)\eta
+
o(\sqrt{T}\|\eta\|).
\label{eq:responsedef}
\end{equation}
All matrix inequalities below use this real lock-in normalization. Other Fourier-amplitude conventions require the corresponding rescaling of both response and Fisher-information rates.

The measured response-to-noise matrix is
\begin{equation}
\mathsf J_{\rm meas}(\omega)
=
\mathsf R^\T(\omega)
\left[\mathsf S^{\out}(\omega)\right]^+
\mathsf R(\omega),
\label{eq:jmeas}
\end{equation}
where the Moore-Penrose inverse acts on the support of $\mathsf S^{\out}$. For a direction $\vartheta$, the scalar $\vartheta^\T\mathsf J_{\rm meas}(\omega)\vartheta$ is the largest response-to-noise ratio per unit time obtainable from a real linear combination of the measured lock-in components.

Let \(\varrho_{\out,\eta}^{T}(\omega)\) be the output-field state over
\([0,T]\) generated by the local model \eqref{eq:realmodel}.  Its
symmetric-logarithmic-derivative quantum Fisher information matrix with
respect to \(\eta\) at \(\eta=0\) is denoted by
\(F_{\out}^{Q,T}(\omega)\) \cite{Helstrom,Holevo,BraunsteinCaves}.

We define the frequency-resolved output-field QFI rate directionally.
For every real direction
\(\vartheta\in\mathbb R^{2n_p}\), set
\begin{equation}
\overline{\calF}_{\out}^{Q}(\omega;\vartheta)
=
\limsup_{T\rightarrow\infty}
\frac{1}{T}
\vartheta^\T
F_{\out}^{Q,T}(\omega)
\vartheta .
\label{eq:qfirate_directional}
\end{equation}
This definition avoids taking a matrix \(\limsup\), which is not a
canonical operation in the Loewner order.  Whenever the ordinary matrix
limit exists, for example in operator norm,
\begin{equation}
\calF_{\out}^{Q}(\omega)
=
\lim_{T\rightarrow\infty}
\frac{1}{T}
F_{\out}^{Q,T}(\omega),
\label{eq:qfirate_matrix_limit}
\end{equation}
then
\[
\overline{\calF}_{\out}^{Q}(\omega;\vartheta)
=
\vartheta^\T
\calF_{\out}^{Q}(\omega)
\vartheta .
\]
Thus all statements below are first formulated as directional quadratic
inequalities; the corresponding matrix inequalities follow whenever the
matrix QFI-rate limit exists.
The derivation is given in Appendix \ref{app:derivation}.

\begin{theorem}[Output-field data-processing bound]
\label{thm:dataprocessing}
Under the mixing assumption \eqref{eq:mixingassumption}, for any
downstream measurement of the output field producing currents
\(\mathbf I(t)\), and for any weak sinusoidal signal at a fixed nonzero
frequency \(\omega\), the measured response-to-noise matrix satisfies
\begin{equation}
\vartheta^\T
\mathsf R^\T(\omega)
\left[\mathsf S^{\out}(\omega)\right]^+
\mathsf R(\omega)
\vartheta
\leq
\overline{\calF}_{\out}^{Q}(\omega;\vartheta)
\label{eq:main1_directional}
\end{equation}
for every real signal direction \(\vartheta\in\mathbb R^{2n_p}\).
If the matrix limit \eqref{eq:qfirate_matrix_limit} exists, this is
equivalently the Loewner inequality
\begin{equation}
\mathsf R^\T(\omega)
\left[\mathsf S^{\out}(\omega)\right]^+
\mathsf R(\omega)
\preceq
\calF_{\out}^{Q}(\omega).
\label{eq:main1}
\end{equation}
\end{theorem}

Theorem~\ref{thm:dataprocessing} is independent of the chosen detector.
Photon counting, homodyne detection, heterodyne detection, inefficient
detection, and adaptive detection merely correspond to different
classical channels or POVMs applied to the same output field, and none
can increase the quantum Fisher information of that field.

For the dissipative tangent \eqref{eq:generalcouplingmod}, define the
signal-activity matrix
\begin{equation}
(\calA_{\rm sig})_{qr}
=
4\Ree
\sum_\mu
\Tr\left[
M_{\mu q}^\dagger M_{\mu r}\rho_{\ssr}
\right].
\label{eq:activitygeneral}
\end{equation}
For the kinetic modulation \eqref{eq:kineticmod}, this reduces to
\begin{equation}
(\calA_{\rm sig})_{qr}
=
\sum_\mu
b_{\mu q}b_{\mu r}
\Tr\left[
L_\mu^\dagger L_\mu\rho_{\ssr}
\right].
\label{eq:activitykinetic}
\end{equation}
The diagonal element
\(\Tr(L_\mu^\dagger L_\mu\rho_{\ssr})\) is the stationary photon flux,
jump rate, or tunneling rate in channel \(\mu\).

\begin{theorem}[Input-output activity bound]
\label{thm:activity}
For the purely dissipative amplitude modulation
\eqref{eq:generalcouplingmod} satisfying
\eqref{eq:purediss_tangent}, with vacuum Markov inputs and stationary
state \(\rho_{\ssr}\), the output-field quantum Fisher information rate
satisfies, for every fixed nonzero frequency \(\omega\) and every real
signal direction \(\vartheta\),
\begin{equation}
\overline{\calF}_{\out}^{Q}(\omega;\vartheta)
\leq
\vartheta^\T
\left(
\calA_{\rm sig}\otimes \id_{2}
\right)
\vartheta .
\label{eq:main2_directional}
\end{equation}
If the matrix limit \eqref{eq:qfirate_matrix_limit} exists, then
\begin{equation}
\calF_{\out}^{Q}(\omega)
\preceq
\calA_{\rm sig}\otimes \id_{2}.
\label{eq:main2}
\end{equation}
For the kinetic modulation \eqref{eq:kineticmod}, the right-hand side is
the channel activity matrix \eqref{eq:activitykinetic}.
\end{theorem}
The physical intuition behind Theorem~2 is particularly simple for
kinetic modulation.  In this case the signal changes the rate of a
dissipative channel, so information about the signal can be written into
the output only through the activity of that channel.  If the channel is
rarely active, a small change of its coupling has correspondingly few
opportunities to leave a distinguishable imprint on the emitted field,
and the output cannot accumulate a large amount of Fisher information
per unit time.  The stationary channel activity therefore sets the
natural information-injection rate available to such a dissipative
signal.  A large activity provides a larger information budget, although
the output QFI need not attain it.

\begin{corollary}[Exact full-output QFI rate]
\label{cor:exact_full_output_qfi}
Assume in addition that the stationary state is faithful and that the complete collection of Markov output channels is retained in the output state.  For a purely dissipative tangent satisfying Eq.~\eqref{eq:purediss_tangent}, the ordinary output-QFI rate matrix exists and is
\begin{equation}
\calF_{\out}^{Q}(\omega)
=
\calA_{\rm sig}\otimes\id_2,
\qquad \omega\neq0.
\label{eq:exact_full_output_qfi}
\end{equation}
In particular, the upper inequality in Eq.~\eqref{eq:mainboxed} is saturated for the full output under kinetic modulation.
\end{corollary}
 A downstream detector may still lose information, so its measured matrix can be strictly smaller.  The output-QFI evaluation and its frequency-mode normalization are derived in Appendix~\ref{app:exact_qfi}.
Combining Theorems~\ref{thm:dataprocessing} and~\ref{thm:activity}
gives the finite-frequency input-output fluctuation-response inequality
in directional form:
\begin{equation}
{
\vartheta^\T
\mathsf R^\T(\omega)
\left[\mathsf S^{\out}(\omega)\right]^+
\mathsf R(\omega)
\vartheta
\leq
\overline{\calF}_{\out}^{Q}(\omega;\vartheta)
\leq
\vartheta^\T
\left(
\calA_{\rm sig}\otimes\id_2
\right)
\vartheta .
}
\label{eq:mainboxed_directional}
\end{equation}
Since this holds for every \(\vartheta\), it also implies the
detector-facing matrix inequality
\begin{equation}
{
\mathsf R^\T(\omega)
\left[\mathsf S^{\out}(\omega)\right]^+
\mathsf R(\omega)
\preceq
\calA_{\rm sig}\otimes\id_2 .
}
\label{eq:mainboxed_activity}
\end{equation}
When the output-field QFI-rate matrix exists, the sharper chain
\begin{equation}
{
\mathsf R^\T(\omega)
\left[\mathsf S^{\out}(\omega)\right]^+
\mathsf R(\omega)
\preceq
\calF_{\out}^{Q}(\omega)
\preceq
\calA_{\rm sig}\otimes\id_2
}
\label{eq:mainboxed}
\end{equation}
is recovered.

In the usual scalar complex notation for one measured current and one
signal direction, the same statement is written as
\begin{equation}
{
\frac{|R_{\rm cplx}(\omega)|^2}{S^{\out}(\omega)}
\leq
\overline{\calF}_{\out}^{Q,{\rm scalar}}(\omega)
\leq
\calA_{\rm sig},
}
\label{eq:mainscalar}
\end{equation}
where \(R_{\rm cplx}\) packages the two real lock-in response components
and
\(\overline{\calF}_{\out}^{Q,{\rm scalar}}(\omega)\) is the corresponding
directional QFI-rate upper limit in the real two-quadrature convention.

There is a closely related coherent-input version. Suppose the signal is a weak coherent displacement of an incoming vacuum field,
\begin{equation}
b_{\mu,\inn}(t)\rightarrow b_{\mu,\inn}(t)+\epsilon f(t),
\label{eq:coherentinput}
\end{equation}
rather than a modulation of $L_\mu$. The input coherent state already carries a quantum Fisher information rate equal to $4$ for the real displacement amplitude in the normalization used here (details given in Appendix \ref{app:coherentbound}). The joint input-system-output evolution is unitary, and tracing out unobserved degrees of freedom cannot increase quantum Fisher information. Hence
\begin{equation}
\mathsf R^\T(\omega)
\left[\mathsf S^{\out}(\omega)\right]^+
\mathsf R(\omega)
\preceq
4\id_2,
\label{eq:coherentdrivebound}
\end{equation}
or in scalar form
\begin{equation}
{
\frac{|R_{\rm cplx}(\omega)|^2}{S^{\out}(\omega)}
\leq
4.
}
\label{eq:mainscalar2}
\end{equation}

This version is useful when the experimentally applied signal is an incoming microwave or optical tone rather than a modulated decay rate.

\section{Remarks on the main results}
\label{sec:remarks}

The conditional matrix chain \eqref{eq:mainboxed}, and its unconditional directional form \eqref{eq:mainboxed_directional}, are not operator fluctuation-dissipation theorems. It does not assert that a retarded commutator is bounded by a symmetrized internal spectrum of the same operator. Instead, it refers to detector records. The matrix $\mathsf S^{\out}(\omega)$ is the covariance matrix of the real lock-in components of currents actually measured outside the system. The response matrix $\mathsf R(\omega)$ is obtained by applying a weak sinusoidal signal and measuring the coherent lock-in output at the same frequency. Both quantities can be obtained without reconstructing the system density matrix.

The inequality is not tied to one unraveling. Photon counting, homodyne detection, heterodyne detection, and adaptive measurements are different measurements on the same output field. The first inequality in Eq. \eqref{eq:mainboxed} says that every such measurement produces a classical Fisher information no larger than the quantum Fisher information of the output field. A fixed quantum-jump trajectory gives a valid classical record, but it is only one possible readout.

The third point concerns the role of noncommutativity. Noncommuting system operators enter through the Lindblad evolution and through the input-output relation. In a driven qubit, for example, a homodyne current measures a field quadrature proportional to a dipole quadrature of the atom. Its spectrum contains Rabi oscillations and phase-sensitive correlations. These effects are absent in a classical jump process with the same average jump rate. The bound is therefore not purely classical, even though the final measured current is a classical time series.

The input-output formulation is essential for experimental interpretation.  Internal system observables are not always directly measured.  Detectors see fields leaving the system, and the detected current contains both system radiation and input noise.  The relation \(b_{\out}=b_{\inn}+L\) provides the link between Lindblad dynamics and observed spectra \cite{GardinerCollett1985,GardinerZoller,ClerkRMP}.  This is why Eq.~\eqref{eq:mainboxed} is formulated in terms of \(\mathsf R(\omega)\) and \(\mathsf S^{\out}(\omega)\), rather than in terms of abstract internal correlation functions.  It is also why homodyne or heterodyne spectra can display coherent quantum features, such as those in resonance fluorescence \cite{Mollow1969}, while still being constrained by a channel activity.  The distinctive feature of the present inequality is therefore not frequency dependence alone, but the placement of an output-field QFI between experimentally measured spectra and a calibrated signal activity.
The right-hand side is deliberately chosen to be experimentally calibratable. For kinetic coupling modulation, $\calA_{\rm sig}$ is built from steady channel fluxes. In a fluorescence experiment this is the photon emission rate. In a transport experiment this is a tunneling rate. In a lossy cavity this is the photon loss flux through a specified port. This is less abstract than a Kubo-Mori or symmetric-logarithmic-derivative correlation spectrum of an internal system operator.



\section{Examples}
\label{sec:examples}

\subsection{Single-sided cavity}

As a preliminary Gaussian benchmark, we consider a passive single-sided cavity. Since its Hilbert space is infinite-dimensional, this example is used only for the coherent-input bound \eqref{eq:coherentdrivebound}, not for the finite-dimensional activity theorem.
Consider a passive single-port cavity with annihilation operator $a$, Hamiltonian
\begin{equation}
H=\Delta a^\dagger a,
\end{equation}
and input-output coupling
\begin{equation}
L=\sqrt{\kappa}\,a .
\end{equation}
The Langevin equation is
\begin{equation}
\dot a(t)
=
-\left(\frac{\kappa}{2}+\ii\Delta\right)a(t)
-\sqrt{\kappa}\,b_{\inn}(t),
\label{eq:cavitylangevin}
\end{equation}
and, with the convention of Eq.~\eqref{eq:inputoutput},
\begin{equation}
b_{\out}(t)=b_{\inn}(t)+\sqrt{\kappa}\,a(t).
\end{equation}
Equation~\eqref{eq:cavitylangevin} is the standard single-port
Markovian input-output Langevin equation \cite{GardinerCollett1985},
obtained by applying the Heisenberg equation with coupling operator
\(L=\sqrt{\kappa}a\).  Solving it in the frequency domain gives
\begin{equation}
a(\omega)
=
-\sqrt{\kappa}\,
\chi_c(\omega)b_{\inn}(\omega),
\qquad
\chi_c(\omega)
=
\frac{1}{\frac{\kappa}{2}+\ii(\Delta-\omega)}.
\end{equation}
Therefore
\begin{equation}
b_{\out}(\omega)
=
s(\omega)b_{\inn}(\omega),
\qquad
s(\omega)=1-\kappa\chi_c(\omega).
\label{eq:cavityscattering}
\end{equation}
For a lossless single-sided cavity,
\begin{equation}
s(\omega)
=
\frac{-\frac{\kappa}{2}+\ii(\Delta-\omega)}
{\frac{\kappa}{2}+\ii(\Delta-\omega)},
\qquad
|s(\omega)|=1 .
\end{equation}

For an ideal homodyne measurement of the output quadrature,
\begin{equation}
I_\theta(t)
=
e^{-\ii\theta}b_{\out}(t)
+
e^{\ii\theta}b_{\out}^\dagger(t),
\end{equation}
the unperturbed input is vacuum.  Since the single-port scattering
relation is passive and lossless, \(|s(\omega)|=1\), the output
quadrature is again a vacuum quadrature.  In the real lock-in
normalization of Eq.~\eqref{eq:realoutspectrum}, its shot-noise spectrum is
therefore
\begin{equation}
S^{\out}(\omega)=1 .
\end{equation}

Now consider a weak coherent displacement of the incoming vacuum field,
\begin{equation}
b_{\inn}(t)\mapsto b_{\inn}(t)+\epsilon f(t).
\end{equation}
In the scalar coherent-input notation of Eq.~\eqref{eq:mainscalar2},
\(R_{\rm cplx}\) is the coefficient of the linear change in the
phase-matched homodyne mean.  The displacement is scattered as
\begin{equation}
\delta\langle b_{\out}(\omega)\rangle
=
s(\omega)\epsilon f(\omega).
\end{equation}
Hence
\begin{equation}
\delta\langle I_\theta(\omega)\rangle
=
2\epsilon f(\omega)\,
\Ree\!\left[e^{-\ii\theta}s(\omega)\right].
\end{equation}
Optimizing over the homodyne phase, \(\theta=\arg s(\omega)\), gives
\begin{equation}
|R_{\rm cplx}(\omega)|
=
\max_\theta
2\left|\Ree\!\left[e^{-\ii\theta}s(\omega)\right]\right|
=
2|s(\omega)|.
\end{equation}
For the lossless single-sided cavity, \(|s(\omega)|=1\).  Therefore
\begin{equation}
\frac{|R_{\rm cplx}(\omega)|^2}{S^{\out}(\omega)}
=
4,
\label{eq:cavity_scalar_bound}
\end{equation}
which saturates the coherent-input data-processing bound \eqref{eq:mainscalar2}.  The cavity
can reshape and phase-shift the signal, but it cannot create information
about a displacement that entered only through the input field.

\subsection{Resonance fluorescence}

Let $\sigma_x$, $\sigma_y$, and $\sigma_z$ be Pauli operators, and let
\begin{equation}
\sigma_-=|g\rangle\langle e|,
\qquad
\sigma_+=\sigma_-^\dagger .
\end{equation}
We consider a driven two-level atom with one measured radiative channel,
\begin{equation}
H=
\frac{\Delta}{2}\sigma_z
+
\frac{\Omega}{2}\sigma_x,
\qquad
L=\sqrt{\kappa}\,\sigma_- .
\label{eq:qubitmodel}
\end{equation}
The output field and ideal homodyne current are
\begin{equation}
b_{\out}(t)=b_{\inn}(t)+\sqrt{\kappa}\,\sigma_-(t),
\qquad
I_\theta(t)
=
e^{-\ii\theta}b_{\out}(t)
+
e^{\ii\theta}b_{\out}^\dagger(t).
\end{equation}
The corresponding system quadrature is
\begin{equation}
X_\theta
=
e^{-\ii\theta}L+e^{\ii\theta}L^\dagger
=
\sqrt{\kappa}\left(\cos\theta\,\sigma_x-\sin\theta\,\sigma_y\right).
\end{equation}

We modulate the radiative coupling amplitude as
\begin{equation}
L^{(\epsilon)}(t)
=
e^{\epsilon(t)/2}L .
\label{eq:qubitmod}
\end{equation}
Since this is the kinetic modulation of Eq.~\eqref{eq:kineticmod}, the activity appearing on the right-hand side of the main bound is the steady fluorescence flux,
\begin{equation}
\calA
=
\Tr[L^\dagger L\rho_{\ssr}]
=
\kappa\langle\sigma_+\sigma_-\rangle_{\ssr}.
\label{eq:qubitactivity}
\end{equation}
For \(\Omega\neq0\), the stationary state is faithful, and the single radiative channel is the complete Markov output.  Corollary~\ref{cor:exact_full_output_qfi} therefore gives the output-field QFI rate explicitly:
\begin{equation}
\calF_{\out}^{Q}(\omega)=\calA\id_2,
\qquad
\calF_{\out}^{Q,\mathrm{scalar}}(\omega)=\calA .
\label{eq:rf_exact_qfi}
\end{equation}
Thus this example displays both levels of the hierarchy,
\begin{equation}
\calJ_\theta(\omega)
\equiv
\frac{|R_\theta(\omega)|^2}{S_\theta^{\out}(\omega)}
\leq
\calF_{\out}^{Q,\mathrm{scalar}}(\omega)
=
\calA .
\label{eq:rf_full_hierarchy}
\end{equation}

We now verify the main inequality explicitly for the resonant case $\Delta=0$. Define
\begin{equation}
D=\kappa^2+2\Omega^2,
\qquad
p_e=\langle\sigma_+\sigma_-\rangle_{\ssr}
=
\frac{\Omega^2}{D},
\qquad
\calA=\kappa p_e .
\end{equation}
The steady-state Bloch components are
\begin{equation}
\langle\sigma_x\rangle_{\ssr}=0,
\qquad
\langle\sigma_y\rangle_{\ssr}=\frac{2\Omega\kappa}{D},
\qquad
\langle\sigma_z\rangle_{\ssr}=-\frac{\kappa^2}{D}.
\end{equation}
With the real unit-RMS lock-in convention of Sec.~\ref{sec:setup}, the coupling modulation produces no linear response in the $\sigma_x$ quadrature. Therefore
\begin{equation}
R_\theta(\omega)
=
-\sin\theta\,R_y(\omega),
\qquad
S_\theta^{\out}(\omega)
=
\cos^2\theta\,S_x(\omega)
+
\sin^2\theta\,S_y(\omega).
\label{eq:rf_phase_split}
\end{equation}
Let
\begin{equation}
d(\omega)=
\left(\frac{\kappa}{2}-\ii\omega\right)
\left(\kappa-\ii\omega\right)
+\Omega^2 .
\end{equation}
The phase-quadrature response is
\begin{equation}
R_y(\omega)
=
\frac{\sqrt{\kappa}\,\Omega\kappa}{D}
\frac{
3\Omega^2-\frac{\kappa^2}{2}-\omega^2-\frac{\ii\kappa\omega}{2}
}{d(\omega)} .
\label{eq:rf_Ry_explicit}
\end{equation}
This expression contains both the dynamical response of the atom and the direct input-output contribution from the explicit dependence of $L^{(\epsilon)}$ on $\epsilon$. The two quadrature spectra are
\begin{equation}
S_x(\omega)
=
1+
\frac{2\kappa^2\Omega^2}
{D\left[(\kappa/2)^2+\omega^2\right]},
\end{equation}
and
\begin{equation}
S_y(\omega)
=
1-
\frac{4\kappa\Omega^2}{D^2}
\Ree
\frac{
\kappa(\kappa^2-4\Omega^2)-\ii\omega(\kappa^2-2\Omega^2)
}{d(\omega)} .
\label{eq:rf_Sy_explicit}
\end{equation}
The leading term $1$ in each spectrum is the vacuum shot-noise contribution.

\begin{figure}[t]
\centering
\includegraphics[width=0.86\columnwidth]{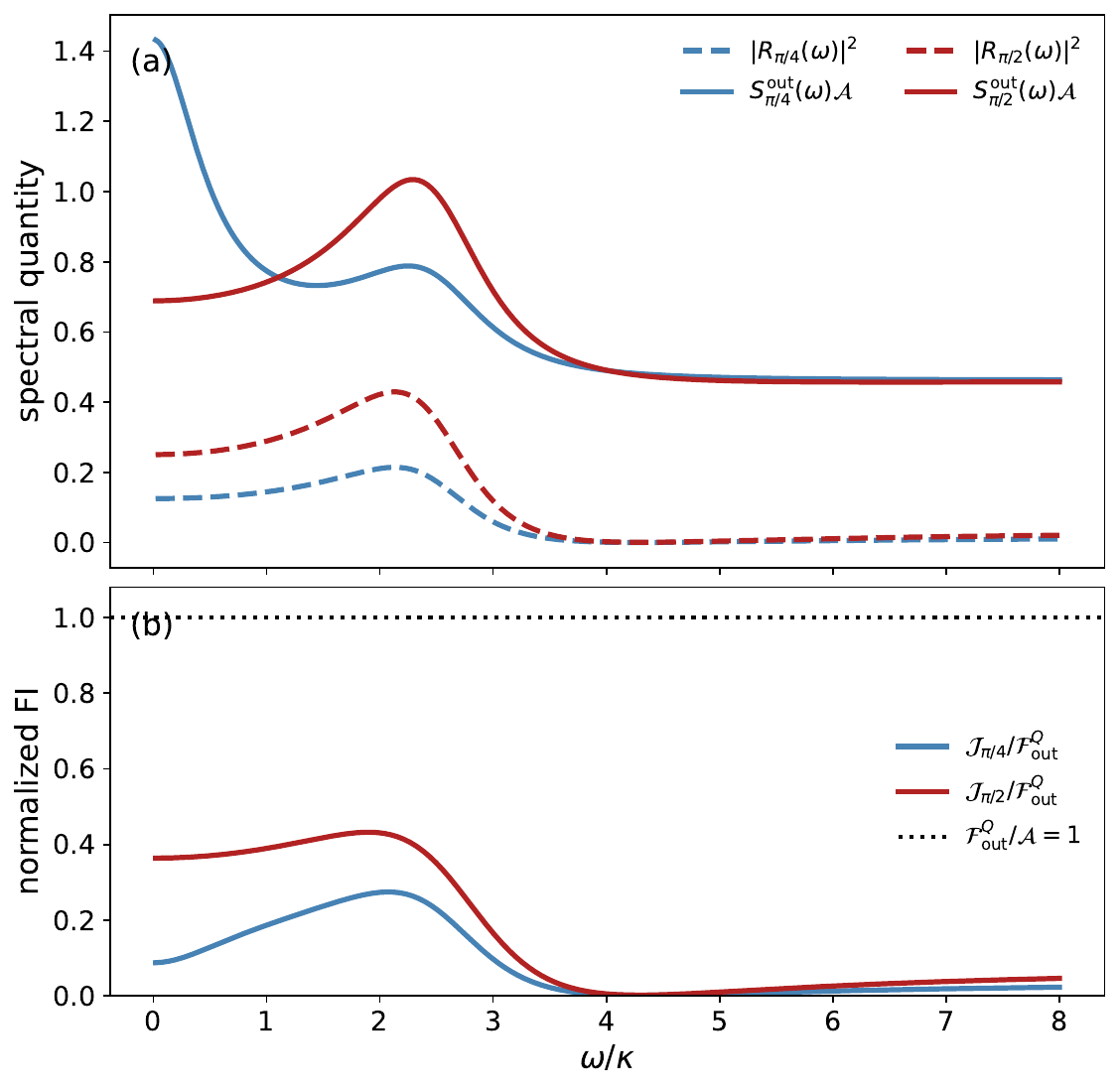}
\caption{
Resonance-fluorescence hierarchy for kinetic modulation of the radiative coupling, with $\Delta=0$ and $\Omega=2.5\kappa$.  (a) The dashed curves are the squared lock-in responses $|R_\theta(\omega)|^2$, and the solid curves are the activity-weighted output spectra $S_\theta^{\out}(\omega)\calA$, for $\theta=\pi/4$ and $\theta=\pi/2$.  The dashed curves remain below the corresponding solid curves, as required by $|R_\theta(\omega)|^2\leq S_\theta^{\out}(\omega)\calA$.  (b) Detector Fisher-information rates $\calJ_\theta=|R_\theta|^2/S_\theta^{\out}$ normalized by the exact full-output QFI rate.  The dotted line is $\calF_{\out}^{Q,\mathrm{scalar}}/\calA=1$, showing that the QFI/activity inequality is saturated by the full output while homodyne detection is generally strict.
}
\label{fig:qubit_homodyne_fri}
\end{figure}

Since $S_x(\omega)>0$, Eq.~\eqref{eq:rf_phase_split} implies
\begin{equation}
\frac{|R_\theta(\omega)|^2}{S_\theta^{\out}(\omega)}
=
\frac{\sin^2\theta\,|R_y(\omega)|^2}
{\cos^2\theta\,S_x(\omega)+\sin^2\theta\,S_y(\omega)}
\leq
\frac{|R_y(\omega)|^2}{S_y(\omega)} ,
\end{equation}
with equality at the phase quadrature $\theta=\pi/2$ whenever $\Omega\neq0$. Direct substitution of Eqs.~\eqref{eq:rf_Ry_explicit} and~\eqref{eq:rf_Sy_explicit} gives
\begin{equation}
\calA S_y(\omega)-|R_y(\omega)|^2
=
\frac{\kappa\Omega^4}
{2D^2|d(\omega)|^2}
\left[
4(\omega^2-\Omega^2)^2
+
\kappa^2(4\Omega^2+9\omega^2+5\kappa^2)
\right]
\geq0 .
\label{eq:rf_explicit_positive}
\end{equation}
Thus
\begin{equation}
\max_\theta
\frac{|R_\theta(\omega)|^2}{S_\theta^{\out}(\omega)}
=
\frac{|R_y(\omega)|^2}{S_y(\omega)}
\leq
\calA ,
\end{equation}
which verifies the detector-level inequality in an exactly solvable finite-dimensional example.  Together with Eq.~\eqref{eq:rf_exact_qfi}, it verifies the complete hierarchy rather than only the combined activity bound.  The undriven case $\Omega=0$ is trivial: the activity and the linear response both vanish.

Figure~\ref{fig:qubit_homodyne_fri} shows the hierarchy numerically for $\Delta=0$ and $\Omega=2.5\kappa$.  In panel (a), the dashed response curves remain below the corresponding solid activity-weighted spectra, correcting the line-style assignment.  Panel (b) compares the detector information directly with the exact output-field QFI.  Over the sampled frequency range, the maximal ratios $\calJ_\theta/\calF_{\out}^{Q,\mathrm{scalar}}$ are $0.275$ for $\theta=\pi/4$ and $0.433$ for $\theta=\pi/2$, whereas $\calF_{\out}^{Q,\mathrm{scalar}}/\calA=1$ identically.

\subsection{Two-port Kerr-parametric cat resonator}
\label{sec:kpo_cat_example}

We next consider a nonlinear bosonic model motivated by Kerr-cat and dissipatively stabilized cat-resonator platforms in circuit QED \cite{PuriBoutinBlais2017,Grimm2020KerrCat,Lescanne2020CatQubit,Reglade2024CatQubit}.  The physical oscillator is infinite-dimensional.  To keep the example within the finite-dimensional assumptions of the main results, all numerical quantities below are defined in a Fock truncation.

Let
\begin{equation}
\calH_N
=
\operatorname{span}\{\ket{0},\ket{1},\ldots,\ket{N-1}\},
\qquad
P_N=\sum_{n=0}^{N-1}\ket n\bra n,
\qquad
a_N=P_NaP_N .
\end{equation}
On \(\calH_N\) we use
\begin{equation}
H_N
=
-\Delta a_N^\dagger a_N
-K(a_N^\dagger)^2a_N^2
+\frac{p}{2}\left[(a_N^\dagger)^2+a_N^2\right]
+F(a_N+a_N^\dagger),
\label{eq:kpo_hamiltonian}
\end{equation}
with two independently structured output couplings,
\begin{equation}
L_{{\rm ex},N}=\sqrt{\kappa_{\rm ex}}\,a_N,
\qquad
L_{2,N}=\sqrt{\kappa_2}\,a_N^2 .
\label{eq:kpo_jumps}
\end{equation}
The first port is a single-photon external-loss channel and the second is an engineered two-photon-loss channel.  Unlike two proportional single-photon couplings, these operators generate independent signal directions and allow appreciable channel mixing.  The two-photon drive \(p\), Kerr nonlinearity \(K\), and small linear bias \(F\) produce a nonlinear cat-resonator regime with a fixed phase-space orientation.  For the parameters below, the truncated generator is mixing and its stationary state \(\rho_{\ssr,N}\) is faithful.

Both output couplings are kinetically modulated,
\begin{equation}
L_{q,N}^{(\epsilon)}(t)
=
e^{\epsilon_q(t)/2}L_{q,N},
\qquad
q\in\{{\rm ex},2\},
\label{eq:kpo_kinetic_mod}
\end{equation}
so that
\begin{equation}
\calV_{q,N}\rho=\calD[L_{q,N}]\rho .
\end{equation}
The signal-activity matrix is diagonal,
\begin{equation}
\calA_{{\rm sig},N}
=
\operatorname{diag}(\calA_{{\rm ex},N},\calA_{2,N}),
\qquad
\calA_{q,N}
=
\Tr[L_{q,N}^\dagger L_{q,N}\rho_{\ssr,N}] .
\label{eq:kpo_activity_matrix}
\end{equation}
Because both Markov outputs are retained, Corollary~\ref{cor:exact_full_output_qfi} gives the intermediate matrix in closed form,
\begin{equation}
\mathsf F_{\out,N}^{Q}(\omega)
=
\calA_{{\rm sig},N}\otimes\id_2 .
\label{eq:kpo_exact_qfi}
\end{equation}

We homodyne both output ports, with phases \(\theta_\alpha\), \(\alpha\in\{{\rm ex},2\}\), and define
\begin{equation}
X_{\alpha,N}
=
e^{-\ii\theta_\alpha}L_{\alpha,N}
+e^{\ii\theta_\alpha}L_{\alpha,N}^\dagger,
\qquad
\calB_{\alpha,N}\rho
=
e^{-\ii\theta_\alpha}L_{\alpha,N}\rho
+e^{\ii\theta_\alpha}\rho L_{\alpha,N}^\dagger .
\label{eq:kpo_Xalpha}
\end{equation}
Writing \(\calQ_NY=Y-\rho_{\ssr,N}\Tr Y\), the complex positive-frequency spectrum matrix is
\begin{align}
C_{\alpha\beta,N}^{\out}(\omega)
&=
\delta_{\alpha\beta}
+
\Tr\!\left[
X_{\alpha,N}(-\ii\omega-\calL_N)^{-1}
\calQ_N(\calB_{\beta,N}\rho_{\ssr,N})
\right]
\nonumber\\
&\quad+
\Tr\!\left[
X_{\beta,N}(\ii\omega-\calL_N)^{-1}
\calQ_N(\calB_{\alpha,N}\rho_{\ssr,N})
\right] .
\label{eq:kpo_complex_spectrum_matrix}
\end{align}
The complex response of output \(\alpha\) to signal \(q\) is
\begin{equation}
\chi_{\alpha q,N}(\omega)
=
\Tr\!\left[
X_{\alpha,N}(-\ii\omega-\calL_N)^{-1}
\calV_{q,N}\rho_{\ssr,N}
\right]
+
\frac{1}{2}\delta_{\alpha q}
\Tr[X_{\alpha,N}\rho_{\ssr,N}] .
\label{eq:kpo_complex_response}
\end{equation}
The direct term occurs when the signal modulates the same coupling operator that appears explicitly in the monitored output quadrature.  Appendix~\ref{app:liouvillian_formulae} derives Eqs.~\eqref{eq:kpo_complex_spectrum_matrix} and \eqref{eq:kpo_complex_response}.

For a complex number \(z\), define the real lock-in block
\begin{equation}
\mathsf B(z)
=
\begin{pmatrix}
\Ree z&-\Imm z\\
\Imm z&\Ree z
\end{pmatrix}.
\label{eq:kpo_real_block}
\end{equation}
In the ordering \((\mathrm{ex},c),(\mathrm{ex},s),(2,c),(2,s)\), the complete real matrices are
\begin{equation}
\mathsf R_N(\omega)
=
\left[\mathsf B(\chi_{\alpha q,N}(\omega))\right]_{\alpha,q},
\qquad
\mathsf S_N^{\out}(\omega)
=
\left[\mathsf B(C_{\alpha\beta,N}^{\out}(\omega))\right]_{\alpha,\beta} .
\label{eq:kpo_real_matrices}
\end{equation}
Thus both \(\mathsf R_N\) and \(\mathsf S_N^{\out}\) are \(4\times4\), and
\begin{equation}
\mathsf J_N(\omega)
=
\mathsf R_N^{\T}(\omega)
[\mathsf S_N^{\out}(\omega)]^{-1}
\mathsf R_N(\omega)
\preceq
\mathsf F_{\out,N}^{Q}(\omega)
=
\calA_{{\rm sig},N}\otimes\id_2 .
\label{eq:kpo_full_hierarchy}
\end{equation}
This explicitly evaluates the intermediate output-field QFI rather than testing only the combined detector/activity bound.

To expose the multiparameter structure, set
\begin{equation}
\widetilde{\mathsf J}_N
=
[\mathsf F_{\out,N}^{Q}]^{-1/2}
\mathsf J_N
[\mathsf F_{\out,N}^{Q}]^{-1/2} .
\label{eq:kpo_normalized_J}
\end{equation}
The complex-to-real structure gives two distinct eigenvalues \(\lambda_+(\omega)\geq\lambda_-(\omega)\), each repeated twice.  For its \(2\times2\) channel blocks, define
\begin{equation}
p_q(\omega)
=
\frac{1}{2}\Tr\widetilde{\mathsf J}_{qq}(\omega),
\qquad
\mu(\omega)
=
\frac{\|\widetilde{\mathsf J}_{{\rm ex},2}(\omega)\|_{\rm F}}{\sqrt2} .
\label{eq:kpo_block_diagnostics}
\end{equation}
Here \(p_q\) measures the normalized diagonal information in channel \(q\), while \(\mu\) measures the off-diagonal coupling between the two signal channels.

For \(N=12\), \(K=1\), \(\Delta=0.2\), \(p=2.0\), \(F=0.15\), \(\kappa_{\rm ex}=\kappa_2=0.2\), and \(\theta_{\rm ex}=\theta_2=0\), the stationary activities are
\begin{equation}
\calA_{{\rm ex},N}=0.1866725,
\qquad
\calA_{2,N}=0.1843478 .
\end{equation}
At the frequency where the mixing diagnostic is maximal, \(\omega_\star/K=3.1567\), the normalized measured-information matrix is
\begin{equation}
\widetilde{\mathsf J}_N(\omega_\star)
=
\begin{pmatrix}
0.661&0&-0.277&-0.024\\
0&0.661&0.024&-0.277\\
-0.277&0.024&0.472&0\\
-0.024&-0.277&0&0.472
\end{pmatrix}.
\label{eq:kpo_numeric_matrix}
\end{equation}
Consequently,
\begin{equation}
\lambda_+=0.860,
\quad
\lambda_-=0.273,
\quad
p_{\rm ex}=0.661,
\quad
p_2=0.472,
\quad
\mu=0.278 .
\label{eq:kpo_numeric_diagnostics}
\end{equation}
Both diagonal channel blocks are appreciable, and the off-diagonal block raises \(\lambda_+\) well above either scalar channel projection.  Over the full displayed interval, \(\max_\omega\lambda_+(\omega)=0.894<1\), while the exact normalized output QFI is unity.

\begin{figure}[t]
\centering
\includegraphics[width=0.95\columnwidth]{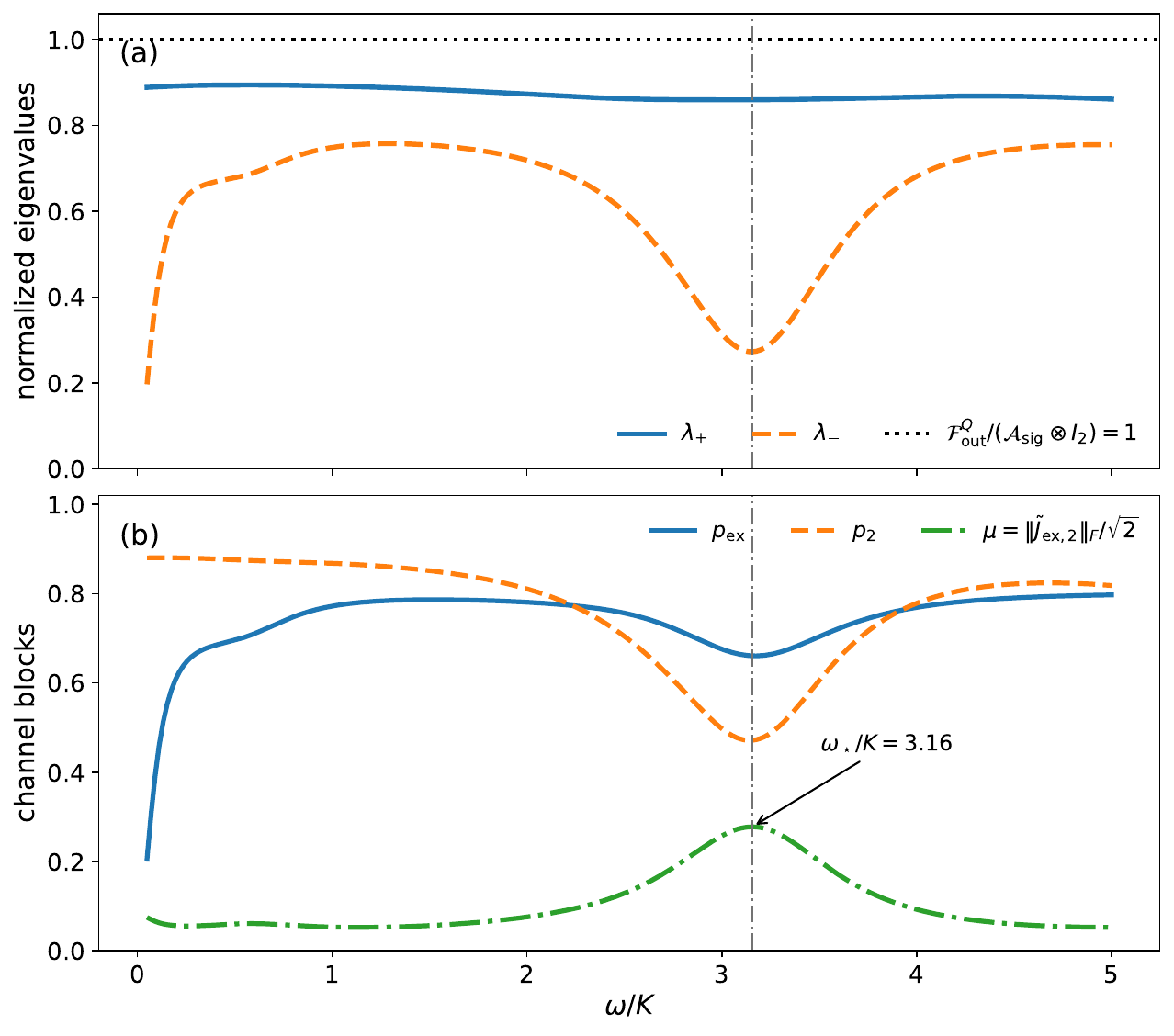}
\caption{
Full two-channel hierarchy for the truncated Kerr-parametric cat resonator.  (a) The two distinct eigenvalues $\lambda_+$ (solid) and $\lambda_-$ (dashed) of $\widetilde{\mathsf J}_N$.  The dotted unity line is the exactly evaluated normalized output-field QFI, $\mathsf F_{\out,N}^{Q}/(\calA_{{\rm sig},N}\otimes\id_2)=1$.  (b) The normalized diagonal channel projections $p_{\rm ex}$ and $p_2$, together with the off-diagonal mixing norm $\mu$.  The dash-dotted vertical line marks $\omega_\star/K=3.16$, where $p_{\rm ex}=0.661$, $p_2=0.472$, and $\mu=0.278$.  Parameters are $N=12$, $K=1$, $\Delta=0.2$, $p=2.0$, $F=0.15$, $\kappa_{\rm ex}=\kappa_2=0.2$, and $\theta_{\rm ex}=\theta_2=0$.
}
\label{fig:kpo_cat_matrix_fri}
\end{figure}

Fig.~\ref{fig:kpo_cat_matrix_fri} summarizes the frequency dependence of the full two-channel hierarchy.  In panel (a), both normalized eigenvalues of $\widetilde{\mathsf J}_N$ remain below the exact full-output QFI ceiling at unity throughout the displayed frequency range, with $\max_\omega \lambda_+(\omega)=0.894$.  The separation between $\lambda_+$ and $\lambda_-$ becomes particularly pronounced near $\omega_\star/K\simeq 3.16$.  Panel (b) resolves this behavior into the two diagonal channel projections and the off-diagonal mixing contribution.  Both $p_{\rm ex}$ and $p_2$ remain appreciable, while the mixing norm $\mu$ develops a clear maximum near $\omega_\star$, where $p_{\rm ex}=0.661$, $p_2=0.472$, and $\mu=0.278$.  At the same frequency, $\lambda_+=0.860$ exceeds either individual channel projection, demonstrating that the off-diagonal block makes a quantitatively significant contribution to the optimal joint signal direction.  Thus the figure illustrates not only the strict detector-to-output-QFI inequality, but also the genuinely multiparameter character of the measured-information matrix.

\section{Relation to existing fluctuation-response bounds}
\label{sec:connection}

This section clarifies the relation between Eq.~\eqref{eq:mainboxed} and existing fluctuation-response bounds.  The equilibrium fluctuation-dissipation theorem relates linear response to equilibrium correlation functions, while nonequilibrium extensions express response in terms of both entropic and frenetic contributions \cite{CallenWelton1951,Kubo1957,Kubo1966,MarconiPuglisiRondoniVulpiani2008,BaiesiMaesWynants2009,SeifertSpeck2010}.  At the spectral level, Harada--Sasa-type relations connect violations of equilibrium fluctuation-response structure to dissipation \cite{HaradaSasa2005,HaradaSasa2006,GuFDRI2026}.  More recent finite-frequency fluctuation-response inequalities instead bound response precision directly by kinetic or activity-like costs, including for Markov jump and Langevin dynamics \cite{DechantPSD2023,Dechant2025,OwenGingrichHorowitz2020,FernandesMartinsHorowitz2023,GaoChunHorowitz2024,AslyamovEsposito2024,PtaszynskiAslyamovEsposito2024}.  The present result is the input-output quantum analogue of this finite-frequency response-to-fluctuation structure.

The connection with Dechant's finite-frequency FRI becomes exact in the classical counting limit \cite{Dechant2025}.  Consider Lindblad jump operators
\begin{equation}
        L_{\alpha\beta}
        =
        \sqrt{\Omega_0(\alpha|\beta)}
        \ket{\alpha}\bra{\beta},
        \qquad \alpha\neq\beta ,
\end{equation}
with diagonal stationary state
\begin{equation}
        \rho_{\ssr}
        =
        \sum_\beta p_{\rm st}(\beta)\ket{\beta}\bra{\beta}.
\end{equation}
A classical kinetic perturbation of the transition rates,
\begin{equation}
        \Omega_\epsilon(\alpha|\beta,t)
        =
        \Omega_0(\alpha|\beta)
        \exp\!\left[
        \sum_q \epsilon_q(t)Y_q(\alpha|\beta)
        \right],
\end{equation}
is represented at the Lindblad-amplitude level by
\begin{equation}
        L_{\alpha\beta}^{(\epsilon)}(t)
        =
        \exp\!\left[
        \frac{1}{2}\sum_q \epsilon_q(t)Y_q(\alpha|\beta)
        \right]
        L_{\alpha\beta}.
\end{equation}
Thus \(b_{\alpha\beta,q}=Y_q(\alpha|\beta)\) in Eq.~\eqref{eq:kineticmod}, and the signal activity \eqref{eq:activitykinetic} reduces to
\begin{equation}
        (\calA_{\rm sig})_{qr}
        =
        \sum_{\alpha,\beta}
        Y_q(\alpha|\beta)Y_r(\alpha|\beta)
        \Omega_0(\alpha|\beta)p_{\rm st}(\beta).
\end{equation}
This is the activity matrix entering the Markov-jump version of finite-frequency fluctuation-response bounds \cite{Dechant2025,OwenGingrichHorowitz2020,FernandesMartinsHorowitz2023,AslyamovEsposito2024,PtaszynskiAslyamovEsposito2024}.  If all jump channels are ideally counted, the output record is the classical jump trajectory.  The real lock-in response matrix \(\mathsf R(\omega)\) and output spectrum \(\mathsf S^{\out}(\omega)\) then reduce, after combining cosine and sine components into the usual complex positive-frequency convention, to the classical response \(R(\omega)\) and power spectrum \(S(\omega)\).  When \(S(\omega)\) is nonsingular, the Moore--Penrose inverse in Eq.~\eqref{eq:mainboxed} becomes the ordinary inverse and one obtains
\begin{equation}
        R^\dagger(\omega)S^{-1}(\omega)R(\omega)
        \leq
        A,
        \qquad
        A_{qr}
        =
        \sum_{\alpha,\beta}
        Y_q(\alpha|\beta)Y_r(\alpha|\beta)
        \Omega_0(\alpha|\beta)p_{\rm st}(\beta).
\end{equation}
Thus the classical counting limit of Eq.~\eqref{eq:mainboxed} is not merely analogous to Dechant's result; it reproduces the same finite-frequency FRI in the jump-process setting \cite{Dechant2025}.

The main results are also closely related to response kinetic uncertainty relations \cite{LiuGuRKUR,LiuGuQRKUR}.  In the classical response KUR, the response precision of a trajectory observable in a Markov jump process is bounded by dynamical activity, up to the maximal logarithmic sensitivity of the transition rates \cite{LiuGuRKUR}.  Related classical response bounds refine this structure by emphasizing kinetic perturbations, activity, and finite-frequency response rather than only entropy production \cite{OwenGingrichHorowitz2020,FernandesMartinsHorowitz2023,GaoChunHorowitz2024,AslyamovEsposito2024,PtaszynskiAslyamovEsposito2024}.  In the quantum response KUR, the same information-theoretic backbone is combined with the quantum Cram\'er--Rao bound and an activity structure for continuously monitored Lindblad dynamics; for steady-state jump observables this yields an activity contribution together with a genuinely quantum inter-subspace term \cite{LiuGuQRKUR}.  The present result follows the same hierarchy,
\[
        \text{response precision}
        \leq
        \text{Fisher information}
        \leq
        \text{activity},
\]
but implements it at the level of the emitted field.  The left-hand side is a finite-frequency response-to-noise matrix of measured output currents, the intermediate term is the output-field QFI, and the upper bound is a calibrated signal-channel activity.

For the kinetic modulation
\begin{equation}
        L_\mu^{(\epsilon)}(t)
        =
        e^{b_\mu \epsilon(t)/2}L_\mu ,
\end{equation}
this activity reduces to
\begin{equation}
        \mathcal A_{\rm sig}
        =
        \sum_\mu b_\mu^2
        \operatorname{Tr}
        \!\left[
            L_\mu^\dagger L_\mu \rho_{\rm ss}
        \right],
\end{equation}
or to its multiparameter matrix generalization.  It is the input-output analogue of the activity appearing in response KURs: a weighted rate of signal-bearing quantum emissions \cite{LiuGuRKUR,LiuGuQRKUR}.  If one further specializes to photon counting, zero-frequency or time-integrated observables, and a classical jump limit, Eq.~\eqref{eq:mainboxed} reduces to an R-KUR-type statement.  Conversely, the input-output formulation goes beyond trajectory-level response KURs by remaining explicitly finite-frequency and by treating phase-sensitive measurements, such as homodyne and heterodyne detection, on the same footing as counting measurements.

This distinction is also important relative to quantum-trajectory fluctuation-response bounds \cite{VanVu2025QuantumTrajectory}.  A trajectory-level inequality is applied after a particular unraveling or monitoring scheme has been chosen.  By contrast, Eq.~\eqref{eq:mainboxed} is imposed before choosing the detector: different detection schemes are different POVMs on the same emitted field, and the measured response-to-noise matrix is bounded by the output-field QFI through data processing.  The bound is therefore unraveling-independent at the field level, while still reducing to classical or trajectory-level activity bounds in the appropriate monitored jump limits.

\section{Discussion and conclusion}
\label{sec:discussion}
\label{sec:conclusion}

The main message of Eq.~\eqref{eq:mainboxed} is that finite-frequency response precision in an output measurement is limited before an unraveling is chosen.  The measured side of the inequality is classical spectral data: the response matrix \(\mathsf R(\omega)\) and the output-current spectrum \(\mathsf S^{\out}(\omega)\).  The intermediate object is the QFI rate of the emitted quantum field, which bounds the information available to any downstream detector.  For dissipative amplitude modulation of Markovian coupling channels, this QFI rate is further bounded by the signal activity \(\calA_{\rm sig}\).  The result therefore occupies a middle ground between two extremes.  It is more quantum than applying a classical fluctuation-response inequality to a fixed trajectory record, but more operational than a bound written only in terms of internal noncommutative correlation functions.

The inequality is most useful as a detector-facing diagnostic bound.  A violation of Eq.~\eqref{eq:mainboxed} would point to missing physics or inconsistent calibration.  Possible causes include unaccounted signal paths, non-Markovian filtering, detector nonlinearities, unmodelled gain, or an incorrect normalization of the measured response.

The bound is a linear-response result. The response matrix is defined at $\epsilon=0$, and the noise spectrum is that of the unperturbed steady state. Nonlinear response can be treated by higher-order information inequalities \cite{BaoLiang2025Nonlinear}, but Eq. \eqref{eq:mainboxed} is a first-order finite-frequency statement.

The frequency independence of $\calA_{\rm sig}$ follows from the Markovian nature of the signal channel. A Markovian bath has no memory, and a local change in the coupling injects distinguishability at a rate fixed by the instantaneous activity. If the bath is non-Markovian, or if the signal is filtered before reaching the system, the right-hand side generally becomes frequency dependent.

The examples illustrate different roles of the bound.  The single-sided cavity gives a Gaussian benchmark in which the coherent-input version can be saturated because the output field is related to the input field by a linear scattering transformation.  The resonance-fluorescence example shows how a finite-dimensional quantum emitter can display phase-sensitive homodyne response while still obeying the same output-field information constraint.  The truncated Kerr-parametric cat resonator provides a non-Gaussian matrix validation in a larger Hilbert space, where the response and noise spectra are no longer reducible to a simple passive scattering problem.  Taken together, these examples show that the inequality is not tied to photon counting, classical jump trajectories, or zero-frequency observables.

The tightness of the bound depends strongly on the measurement architecture and on the amount of unobserved information loss.  Linear passive systems driven and measured through the relevant port can saturate the coherent-input version.  Unobserved loss channels, thermal noise, inefficient detection, and internal nonlinear dynamics generally make the inequality strict.  Inefficient detection can be modelled by inserting beam splitters before ideal detectors; this can only decrease the measured signal-to-noise ratio, so the bound remains valid, but the achievable precision moves farther below the activity limit.  Nonlinear quantum systems may approach the bound near resonant features and for optimized quadrature measurements, but exact saturation should not be expected generically.

Several assumptions should be kept explicit.  The derivation assumes Markovian input fields, weak sinusoidal signals, stationary operation, and calibrated accounting of the relevant output channels.  If the bath has memory, the information injected by the signal may become frequency dependent, and a frequency-independent activity bound need not hold in the present form.  If the signal changes the stationary state strongly or drives the system into a nonlinear-response regime, higher-order response terms are required.  If some output channels are unobserved or if the detector chain contains nontrivial filtering, the measured spectra must be interpreted as spectra of the accessible field after those transformations, not as spectra of the ideal output field itself.

The multiparameter case also raises a genuine quantum issue.  The scalar inequality along any fixed signal direction follows from the ordinary single-parameter QFI bound.  For several signal quadratures or several modulated channels, however, the SLD-QFI matrix need not be jointly attainable by a single measurement when the corresponding parameters are incompatible.  The matrix inequality in Eq.~\eqref{eq:mainboxed} should therefore be read as an information upper bound on all downstream measurements, not automatically as a jointly saturable estimation bound.  A finite-frequency Holevo-type formulation is the natural next step when simultaneous optimal estimation of incompatible signal components is required \cite{Holevo}.

Another important extension is the Hamiltonian signal case.  If the perturbation is
\begin{equation}
        H(t)=H_0-\epsilon f(t)B ,
\end{equation}
then the signal is not injected through a jump activity.  The corresponding information cost should instead be controlled by a quantum Fisher strength associated with the Hamiltonian tangent vector generated by \(B\).  Such a result would be closer to a noncommutative operator-level FRI and would generally not reduce to a simple channel flux.  The input-output data-processing part of Eq.~\eqref{eq:mainboxed} should still apply, but the simple activity expression \eqref{eq:activitykinetic} is special to dissipative amplitude modulation of Markovian coupling operators.

The most direct experimental settings for testing the bound are platforms where input-output theory and calibrated detection are already standard, including resonance fluorescence, circuit QED, optomechanics, and mesoscopic transport.  Homodyne and heterodyne detection are particularly useful because they access coherent quadrature response that may be invisible in simple counting statistics.  In such settings the inequality provides a direct consistency relation between measured finite-frequency signal-to-noise ratio and independently calibrated signal-channel activity.

In summary, we have formulated a finite-frequency fluctuation-response inequality for open quantum systems in an input-output setting.  At each frequency, the optimal linear response-to-noise ratio of measured output currents is bounded by the QFI rate of the emitted field, and for dissipative amplitude modulation of Markovian coupling channels this QFI rate is bounded by a signal activity.  For faithful stationary dynamics with the complete Markov output retained, we evaluated the intermediate QFI explicitly for horizontal kinetic tangents and found that it equals the activity matrix. The central structural feature is detector-facing but unraveling-independent: different measurements are treated as different POVMs on the same output field, while the bound itself is imposed before any of them is selected.  Extending this structure to non-Markovian environments, Hamiltonian perturbations, nonlinear response, and multiparameter incompatible estimation would clarify how broadly finite-frequency response precision in quantum open systems is constrained by information injected into the output field.

\appendix

\section{Derivation of the main results}
\label{app:derivation}

\subsection{Real frequency modes and measured spectral signal-to-noise ratio}

The proof begins with an elementary classical fact. Let $Z$ denote the complete measurement record generated by a chosen detector during $[0,T]$, and let $P_\vartheta^T(Z)$ be its probability distribution under a real parameter vector $\vartheta$. For any real statistic $X(Z)$ with mean $m(\vartheta)$ and covariance $\Sigma$, define the score $\ell_i(Z)=\partial_i\ln P_\vartheta^T(Z)|_{\vartheta=0}$. Then
\begin{equation}
\partial_i m
=
\mathbb E\left[(X-m)\ell_i\right].
\label{eq:scoreidentity}
\end{equation}
The block covariance matrix of $(X,\ell)$ is positive semidefinite. Taking its Schur complement gives
\begin{equation}
B^\T\Sigma^+B\preceq F_{\rm cl}^{T},
\label{eq:classicalprojection}
\end{equation}
where $B_{ai}=\partial_i m_a$ and $F_{{\rm cl},ij}^{T}=\mathbb E(\ell_i\ell_j)$ is the classical Fisher information matrix. Equation \eqref{eq:classicalprojection} does not require the statistic to be Gaussian. A central-limit theorem is useful for interpreting the Fourier modes as Gaussian spectral estimators, but the inequality itself is just the score identity plus Cauchy-Schwarz.

We apply Eq. \eqref{eq:classicalprojection} to the real lock-in vector
\begin{equation}
\mathsf X_T(\omega)
=
\left(
\sqrt2\Ree\tilde I_{1,T},\sqrt2\Imm\tilde I_{1,T},\ldots,
\sqrt2\Ree\tilde I_{m,T},\sqrt2\Imm\tilde I_{m,T}
\right)^\T .
\label{eq:realfourierstat}
\end{equation}
The perturbation parameters are the real quadrature amplitudes $\eta=(\eta_{1,c},\eta_{1,s},\ldots,\eta_{n_p,c},\eta_{n_p,s})^\T$ in Eq. \eqref{eq:realmodel}. Stationarity implies that, at fixed nonzero $\omega$, the covariance of $\mathsf X_T(\omega)$ is the real lock-in spectral matrix $\mathsf S^{\out}(\omega)$ defined in Eq.~\eqref{eq:realoutspectrum}. With the unit-RMS convention \eqref{eq:unitrmsmodes}, the Schur complement inequality becomes
\begin{equation}
T
\mathsf R^\T(\omega)
\left[\mathsf S^{\out}(\omega)\right]^+
\mathsf R(\omega)
\preceq
F_{\rm cl}^{T}(\omega)+o(T).
\label{eq:clfibound}
\end{equation}
Equivalently, for a chosen real direction $\vartheta$, maximizing over all real lock-in filters $u$ gives the generalized Rayleigh quotient
\begin{equation}
\max_u
\frac{|u^\T\mathsf R(\omega)\vartheta|^2}
{u^\T\mathsf S^{\out}(\omega)u}
=
\vartheta^\T\mathsf R^\T(\omega)
\left[\mathsf S^{\out}(\omega)\right]^+
\mathsf R(\omega)\vartheta .
\label{eq:appendixschur}
\end{equation}
This is the operational meaning of the measured response-to-noise matrix. If the two real lock-in components are repackaged into one complex positive-frequency variable, the factors of $\sqrt2$ in Eq.~\eqref{eq:realfourierstat} are essential; omitting them is precisely what produces the spurious factor-of-two mismatch between the complex Fourier convention and the real-mode Fisher information.

\subsection{Data processing from the output field to the detector record}

Let \(\varrho_{\out,\vartheta}^{T}\) be the output-field state over
\([0,T]\) in the local model \eqref{eq:realmodel}.  A detector is a POVM
\(\{E_Z\}\) on this field, possibly after adding ancillary vacuum modes
and including classical feedback.  The observed distribution is
\begin{equation}
P_\vartheta^T(Z)
=
\Tr\left[
E_Z\varrho_{\out,\vartheta}^{T}
\right].
\label{eq:povmprob}
\end{equation}
Quantum Fisher information is monotone under completely positive
trace-preserving maps and, in particular, under measurement.  Therefore,
for every finite \(T\),
\begin{equation}
F_{\rm cl}^{T}(\omega)
\preceq
F_{\out}^{Q,T}(\omega).
\label{eq:qfidataprocess}
\end{equation}

Combining the finite-time classical projection inequality
\eqref{eq:clfibound} with \eqref{eq:qfidataprocess} gives, for every
real signal direction \(\vartheta\),
\begin{equation}
T\,
\vartheta^\T
\mathsf R^\T(\omega)
\left[
\mathsf S^{\out}(\omega)
\right]^+
\mathsf R(\omega)
\vartheta
\leq
\vartheta^\T
F_{\out}^{Q,T}(\omega)
\vartheta
+
o(T).
\label{eq:directional_dataprocess_finiteT}
\end{equation}
Dividing by \(T\) and taking the upper long-time limit gives
\begin{equation}
\vartheta^\T
\mathsf R^\T(\omega)
\left[
\mathsf S^{\out}(\omega)
\right]^+
\mathsf R(\omega)
\vartheta
\leq
\overline{\calF}_{\out}^{Q}(\omega;\vartheta).
\end{equation}
This proves Theorem~\ref{thm:dataprocessing} in its directional form.
If \(T^{-1}F_{\out}^{Q,T}(\omega)\) has a matrix limit, the same
inequality for all \(\vartheta\) is equivalent to the Loewner inequality
\eqref{eq:main1}.  Notice that no assumption has been made about which
output measurement is chosen.  The measurement choice enters only
through the POVM in Eq.~\eqref{eq:povmprob}, and data processing removes
it from the upper bound.

\subsection{Continuous-time Stinespring bound for dissipative coupling tangents}

We prove the activity bound directionally.  Fix a real signal direction
\(\vartheta\in\mathbb R^{2n_p}\), with components
\(\vartheta_{q,a}\) where \(a\in\{c,s\}\).  Along this direction the
coupling tangent is
\begin{equation}
N_\mu^\vartheta(t)
=
\sum_{q=1}^{n_p}
\sum_{a\in\{c,s\}}
\vartheta_{q,a}\,
\phi_a(t)\,
M_{\mu q},
\label{eq:directional_tangent}
\end{equation}
so that
\begin{equation}
L_\mu^{(\epsilon\vartheta)}(t)
=
L_\mu+\epsilon N_\mu^\vartheta(t)+O(\epsilon^2).
\end{equation}

Discretize time into bins of length \(\Delta t\).  Over a bin centered at
time \(t_k\), the vacuum Stinespring representation may be written as
\begin{equation}
V_\epsilon(t_k)
=
K_{0,k}^{(\epsilon)}\otimes |0\rangle
+
\sum_\mu
K_{\mu,k}^{(\epsilon)}\otimes |1_\mu\rangle
+
O(\Delta t),
\label{eq:shortisometry_discrete}
\end{equation}
with
\begin{align}
K_{0,k}^{(\epsilon)}
&=
\id
-
\left[
\ii H+
\frac{1}{2}
\sum_\mu
L_\mu^{(\epsilon)\dagger}(t_k)
L_\mu^{(\epsilon)}(t_k)
\right]\Delta t
+
O(\Delta t^2),
\label{eq:k0eps_discrete}
\\
K_{\mu,k}^{(\epsilon)}
&=
L_\mu^{(\epsilon)}(t_k)\sqrt{\Delta t}
+
O(\Delta t^{3/2}).
\label{eq:kmueps_discrete}
\end{align}
The derivatives at \(\epsilon=0\) are
\begin{align}
\dot K_{\mu,k}
&=
N_\mu^\vartheta(t_k)\sqrt{\Delta t}
+
O(\Delta t^{3/2}),
\label{eq:jumpderiv_directional}
\\
\dot K_{0,k}
&=
-\frac{1}{2}
\sum_\mu
\left[
N_\mu^\vartheta(t_k)^\dagger L_\mu
+
L_\mu^\dagger N_\mu^\vartheta(t_k)
\right]\Delta t
+
O(\Delta t^2),
\label{eq:noderiv_directional}
\end{align}
where the dot denotes \(\partial_\epsilon|_{\epsilon=0}\).

The purely dissipative condition \eqref{eq:purediss_tangent} implies the
parallel-tangent relation
\begin{equation}
\sum_i
K_{i,k}^{(0)\dagger}\dot K_{i,k}
=
O(\Delta t^2),
\label{eq:parallel_tangent}
\end{equation}
where \(i=0,\mu\).  This condition removes the coherent
Hamiltonian-like channel tangent that would otherwise accumulate in a
sequential experiment.  Without \eqref{eq:purediss_tangent}, an
additional channel-QFI term appears and the activity-only bound is not
the correct general statement.

Let the pre-bin system state be \(\rho\), and purify it by an arbitrary
reference.  Applying \(V_\epsilon(t_k)\) to the system gives a pure
joint state \(|\Psi_{\epsilon,k}\rangle\) of system, reference, and the
field bin.  The QFI of the field bin after tracing out the system and
reference is no larger than the QFI of this purified joint state.  The
pure-state formula gives
\begin{equation}
F_{k}^{Q}(\rho;\vartheta)
\leq
4
\left(
\langle \dot\Psi_k|\dot\Psi_k\rangle
-
|\langle \Psi_k|\dot\Psi_k\rangle|^2
\right).
\label{eq:local_pure_qfi}
\end{equation}
Using Eqs.~\eqref{eq:jumpderiv_directional}--\eqref{eq:parallel_tangent}
one obtains the local bound
\begin{equation}
F_{k}^{Q}(\rho;\vartheta)
\leq
4
\sum_\mu
\Tr\left[
N_\mu^\vartheta(t_k)^\dagger
N_\mu^\vartheta(t_k)
\rho
\right]\Delta t
+
C\,\Delta t^2,
\label{eq:localqfibound_directional}
\end{equation}
where \(C\) is uniform because the system Hilbert space is finite
dimensional and all operators are bounded.  The no-jump derivative
contributes only to the \(O(\Delta t^2)\) remainder, while the jump
derivatives give the leading \(O(\Delta t)\) term.

For the full time interval, concatenate the bin isometries and retain
the final system and all field bins.  The actual observed output field
is obtained from this enlarged state by tracing out the final system and
any unobserved fields, so monotonicity of QFI allows us to bound the
output-field QFI by the QFI of the enlarged sequential state.  Applying
the amortized channel-QFI chain rule to the sequence of bins gives
\begin{equation}
F_{\out}^{Q,T}(\omega;\vartheta)
\leq
4
\sum_{k}
\sum_\mu
\Tr\left[
N_\mu^\vartheta(t_k)^\dagger
N_\mu^\vartheta(t_k)
\rho_{0}(t_k)
\right]\Delta t
+
O(T\Delta t),
\label{eq:discrete_integrated_qfi}
\end{equation}
where \(\rho_0(t_k)\) is the unperturbed pre-bin system state.  The
\(O(T\Delta t)\) term vanishes in the continuous-time limit at fixed
\(T\).  Since the unperturbed process is initialized in the stationary
state, \(\rho_0(t_k)=\rho_{\ssr}\).  Passing to the continuous-time
limit therefore yields
\begin{equation}
F_{\out}^{Q,T}(\omega;\vartheta)
\leq
4
\int_0^T
\dd t\,
\sum_\mu
\Tr\left[
N_\mu^\vartheta(t)^\dagger
N_\mu^\vartheta(t)
\rho_{\ssr}
\right]
+
o(T).
\label{eq:integratedqfi_directional}
\end{equation}

Substituting the definition \eqref{eq:directional_tangent} gives
\begin{align}
\frac{1}{T}
F_{\out}^{Q,T}(\omega;\vartheta)
&\leq
\sum_{q,r}
\sum_{a,b\in\{c,s\}}
\vartheta_{q,a}\vartheta_{r,b}
\left[
\frac{1}{T}
\int_0^T
\phi_a(t)\phi_b(t)\,\dd t
\right]
4\Ree
\sum_\mu
\Tr\left[
M_{\mu q}^\dagger
M_{\mu r}
\rho_{\ssr}
\right]
+
o(1).
\end{align}
For fixed nonzero \(\omega\), the real unit-RMS modes satisfy
\begin{equation}
\frac{1}{T}
\int_0^T
\phi_a(t)\phi_b(t)\,\dd t
\longrightarrow
\delta_{ab}.
\end{equation}
Taking the upper long-time limit gives
\begin{equation}
\overline{\calF}_{\out}^{Q}(\omega;\vartheta)
\leq
\sum_{q,r}
\sum_{a\in\{c,s\}}
\vartheta_{q,a}\vartheta_{r,a}
(\calA_{\rm sig})_{qr}
=
\vartheta^\T
\left(
\calA_{\rm sig}\otimes\id_2
\right)
\vartheta .
\label{eq:activity_directional_final}
\end{equation}
This proves Theorem~\ref{thm:activity}.  If the matrix QFI-rate limit
\eqref{eq:qfirate_matrix_limit} exists, the same directional inequality
for all \(\vartheta\) is equivalent to
\[
\calF_{\out}^{Q}(\omega)
\preceq
\calA_{\rm sig}\otimes\id_2 .
\]

For the kinetic modulation \(M_{\mu q}=b_{\mu q}L_\mu/2\), the activity
matrix becomes
\begin{equation}
(\calA_{\rm sig})_{qr}
=
\sum_\mu b_{\mu q}b_{\mu r}
\Tr\left[
L_\mu^\dagger L_\mu\rho_{\ssr}
\right],
\end{equation}
which is Eq.~\eqref{eq:activitykinetic}.  The right-hand side is the
steady event flux in the modulated channels, weighted by the calibrated
signal-coupling coefficients.

\subsection{Exact full-output QFI for horizontal dissipative tangents}
\label{app:exact_qfi}

The activity bound above follows from channel monotonicity and therefore does not require an explicit evaluation of the output QFI.  For the full output of a faithful, mixing finite-dimensional quantum Markov model, that evaluation is available from continuous-time quantum Markov local asymptotic normality and output information geometry \cite{CatanaBoutenGuta2015,GutaKiukas2017}.  We summarize the specialization needed here in the standard SLD-QFI normalization.

For parameter \(q\), the dynamical tangent is
\begin{equation}
\dot D_q=(0,M_{1q},\ldots,M_{n_cq}) .
\end{equation}
The quantum-It\^o horizontal component entering the output-QFI metric is
\begin{equation}
\calE_q
=
\Imm\sum_\mu M_{\mu q}^\dagger L_\mu .
\label{eq:qfi_horizontal_component}
\end{equation}
Condition~\eqref{eq:purediss_tangent} is exactly \(\calE_q=0\).  Hence no Poisson-equation or gauge-projection correction is required, and the full-output QFI-rate metric for two such tangents is
\begin{equation}
f_{qr}^{Q}
=
4\Ree\sum_\mu
\Tr[M_{\mu q}^\dagger M_{\mu r}\rho_{\ssr}] .
\label{eq:qfi_horizontal_metric}
\end{equation}
The same rate is obtained for the reduced output state after tracing out the finite-dimensional final system; mixing makes the final-system contribution subextensive, and the system-output and reduced-output local models have the same Gaussian limit \cite{CatanaBoutenGuta2015,GutaKiukas2017}.

For the deterministic frequency envelopes in Eq.~\eqref{eq:unitrmsmodes}, the tangent for component \((q,a)\) is \(\phi_a(t)M_{\mu q}\).  It remains horizontal at every time.  The corresponding infinitesimal generator is a quantum stochastic martingale, so its It\^o covariance is local in time.  Consequently,
\begin{align}
\lim_{T\to\infty}\frac{1}{T}
(F_{\out}^{Q,T})_{(q,a),(r,b)}
&=
4\Ree\sum_\mu
\Tr[M_{\mu q}^\dagger M_{\mu r}\rho_{\ssr}]
\lim_{T\to\infty}\frac{1}{T}
\int_0^T\phi_a(t)\phi_b(t)\dd t
\nonumber\\
&=
(\calA_{\rm sig})_{qr}\,\delta_{ab} .
\label{eq:qfi_frequency_exact}
\end{align}
This proves Eq.~\eqref{eq:exact_full_output_qfi} and, in particular, establishes the ordinary matrix limit needed for the Loewner chain in the two dissipative examples.  For kinetic modulation, substituting \(M_{\mu q}=b_{\mu q}L_\mu/2\) recovers the stationary flux matrix \eqref{eq:activitykinetic} exactly.

\section{Coherent input drive bound}
\label{app:coherentbound}

For a coherent input displacement,
\begin{equation}
        b_{\inn}(t)\rightarrow b_{\inn}(t)+\epsilon f(t),
\end{equation}
the parameter is encoded in the incoming field before the field interacts with the system.  Let
\begin{equation}
        f(t)=u\,\phi_a(t),
        \qquad
        a\in\{c,s\},
\end{equation}
where \(u\) is a fixed input-channel vector and the real frequency modes are normalized as
\begin{equation}
        \frac{1}{T}\int_0^T \phi_a(t)\phi_b(t)\,\dd t
        \longrightarrow
        \delta_{ab}.
\end{equation}
For each real mode define
\begin{equation}
        B_{u,a}
        =
        T^{-1/2}
        \int_0^T
        \phi_a(t)\,u^\dagger\dd B_{\inn}(t),
        \qquad
        [B_{u,a},B_{u,a}^\dagger]=1 ,
\end{equation}
assuming \(u^\dagger u=1\).  The incoming vacuum displaced by a small real amplitude \(\epsilon\) is therefore, in this mode,
\begin{equation}
        \ket{\psi_\epsilon}
        =
        \exp\!\left[
        \epsilon\sqrt{T}
        (B_{u,a}^\dagger-B_{u,a})
        \right]\ket{0}.
\end{equation}
Using the pure-state formula
\begin{equation}
        F_Q
        =
        4\left(
        \braket{\partial_\epsilon\psi_\epsilon|\partial_\epsilon\psi_\epsilon}
        -
        |\braket{\psi_\epsilon|\partial_\epsilon\psi_\epsilon}|^2
        \right),
\end{equation}
one obtains \(F_Q^T=4T\), and hence a quantum Fisher information rate equal to \(4\) for each unit-RMS real input quadrature.  The cross term between the cosine and sine modes vanishes in the long-time limit by their orthogonality, so the input-field QFI rate is \(4\id_2\) in the two-dimensional real quadrature space.

The subsequent system-field evolution is unitary on the joint input, system, and output degrees of freedom.  Discarding the system or unobserved output channels is a quantum channel, and therefore cannot increase quantum Fisher information.  Hence, for every real input-quadrature direction \(\vartheta\),
\begin{equation}
\overline{\calF}_{\out}^{Q}(\omega;\vartheta)
\leq
4\,\vartheta^\T\vartheta .
\end{equation}
If the corresponding matrix QFI-rate limit exists, this is equivalently
\[
\calF_{\out}^{Q}(\omega)\preceq 4\id_2 .
\]
Together with Theorem~\ref{thm:dataprocessing}, this gives
Eq.~\eqref{eq:coherentdrivebound}.

\section{Liouvillian formulae for response and spectra}
\label{app:liouvillian_formulae}

For completeness, we spell out the Liouvillian formulae used in the examples.  The purpose of this subsection is not to introduce an additional assumption, but to fix the precise convention behind the resolvents, the homodyne insertion superoperator, and the direct input-output term.

Let $\calL$ be the unperturbed Lindblad generator,
\begin{equation}
\calL\rho
=
-\ii[H,\rho]+\sum_\mu \calD[L_\mu]\rho ,
\end{equation}
and let $\rho_{\ssr}$ be its stationary state,
\begin{equation}
\calL\rho_{\ssr}=0,
\qquad
\Tr\rho_{\ssr}=1 .
\end{equation}
We use the projection
\begin{equation}
\calQ Y
=
Y-\rho_{\ssr}\Tr Y ,
\label{eq:app_Q_projection}
\end{equation}
which removes the stationary trace component.  On the traceless subspace, the mixing assumption implies that
$ -\ii\omega-\calL
$
is invertible for every real nonzero $\omega$.  Equivalently, throughout this subsection one may read
$(-\ii\omega-\calL)^{-1}$ as the reduced resolvent acting on the $\calQ$-projected subspace.

Suppose first that the system generator is weakly perturbed as
\begin{equation}
\dot\rho(t)
=
\calL\rho(t)+\epsilon(t)\calV\rho(t)+O(\epsilon^2),
\label{eq:app_perturbed_master}
\end{equation}
where $\calV=\partial_\epsilon\calL^{(\epsilon)}|_{\epsilon=0}$ is the first-order perturbation superoperator.  Linearizing around the stationary state,
\begin{equation}
\rho(t)=\rho_{\ssr}+\delta\rho(t)+O(\epsilon^2),
\end{equation}
gives
\begin{equation}
\delta\dot\rho(t)
=
\calL\delta\rho(t)+\epsilon(t)\calV\rho_{\ssr}.
\label{eq:app_linearized_master}
\end{equation}
For a positive-frequency perturbation
\begin{equation}
\epsilon(t)=\epsilon_\omega e^{-\ii\omega t},
\qquad
\delta\rho(t)=\epsilon_\omega e^{-\ii\omega t}\delta\rho(\omega),
\end{equation}
Eq.~\eqref{eq:app_linearized_master} gives
\begin{equation}
(-\ii\omega-\calL)\delta\rho(\omega)
=
\calV\rho_{\ssr}.
\end{equation}
Thus
\begin{equation}
\delta\rho(\omega)
=
(-\ii\omega-\calL)^{-1}\calV\rho_{\ssr}.
\label{eq:app_density_response}
\end{equation}
Since $\Tr[\calV\rho_{\ssr}]=0$ for a trace-preserving perturbation of the generator, the source term lies in the traceless subspace, where the reduced resolvent is well defined.

Now let the measured output current have the form
\begin{equation}
I_a(t)
=
\xi_a(t)+X_a(t),
\label{eq:app_current_decomposition}
\end{equation}
where $\xi_a(t)$ denotes the vacuum input noise contribution and $X_a$ is the system operator appearing in the corresponding output quadrature.  For homodyne detection of one output channel with coupling operator $L$ and local-oscillator phase $\theta$,
\begin{equation}
I_\theta(t)
=
e^{-\ii\theta}b_{\out}(t)+e^{\ii\theta}b_{\out}^\dagger(t)
=
\xi_\theta(t)+X_\theta(t),
\end{equation}
with
\begin{equation}
\xi_\theta(t)
=
e^{-\ii\theta}b_{\inn}(t)+e^{\ii\theta}b_{\inn}^\dagger(t),
\qquad
X_\theta
=
e^{-\ii\theta}L+e^{\ii\theta}L^\dagger .
\label{eq:app_Xtheta_def}
\end{equation}
The response of the mean current has two contributions.  The first is the indirect response caused by the perturbation of the density matrix.  Using Eq.~\eqref{eq:app_density_response}, this contribution is
\begin{equation}
R_{a}^{\rm ind}(\omega)
=
\Tr\!\left[
X_a(-\ii\omega-\calL)^{-1}\calV\rho_{\ssr}
\right].
\label{eq:app_indirect_response}
\end{equation}
The second is a direct response, present only if the measured output operator itself depends explicitly on the signal.  If
\begin{equation}
X_a^{(\epsilon)}
=
X_a+\epsilon\,\dot X_a+O(\epsilon^2),
\qquad
\dot X_a=\partial_\epsilon X_a^{(\epsilon)}|_{\epsilon=0},
\end{equation}
then the direct contribution is
\begin{equation}
R_a^{\rm dir}
=
\Tr[\dot X_a\rho_{\ssr}].
\label{eq:app_direct_response}
\end{equation}
Therefore the full complex response coefficient is
\begin{equation}
R_a(\omega)
=
\Tr\!\left[
X_a(-\ii\omega-\calL)^{-1}\calV\rho_{\ssr}
\right]
+
R_a^{\rm dir}.
\label{eq:app_total_response}
\end{equation}
For several signal parameters $\epsilon_q$, one replaces $\calV$ by $\calV_q$ and $\dot X_a$ by $\dot X_{a,q}$:
\begin{equation}
R_{a,q}(\omega)
=
\Tr\!\left[
X_a(-\ii\omega-\calL)^{-1}\calV_q\rho_{\ssr}
\right]
+
\Tr[\dot X_{a,q}\rho_{\ssr}].
\label{eq:app_multi_response}
\end{equation}

In the two-port truncated cat-resonator example, both monitored coupling operators are modulated according to Eq.~\eqref{eq:kpo_kinetic_mod}.  For output \(\alpha\) and signal \(q\),
\begin{equation}
X_{\alpha,N}^{(\epsilon_q)}
=
X_{\alpha,N}
+
\frac{\epsilon_q}{2}\delta_{\alpha q}X_{\alpha,N}
+
O(\epsilon_q^2),
\end{equation}
so that
\begin{equation}
\dot X_{\alpha,N;q}
=
\frac{1}{2}\delta_{\alpha q}X_{\alpha,N} .
\end{equation}
Equation~\eqref{eq:app_multi_response} therefore gives
\begin{equation*}
\chi_{\alpha q,N}(\omega)
=
\Tr\!\left[
X_{\alpha,N}
(-\ii\omega-\calL_N)^{-1}
\calV_{q,N}\rho_{\ssr,N}
\right]
+
\frac{1}{2}\delta_{\alpha q}
\Tr[X_{\alpha,N}\rho_{\ssr,N}] .
\end{equation*}
This reproduces Eq.~\eqref{eq:kpo_complex_response} and is the response formula used in the complete two-output calculation.

We next derive the homodyne spectrum.  For a stationary current, define the connected correlation function
\begin{equation}
C_{ab}(\tau)
=
\langle \delta I_a(t+\tau)\delta I_b(t)\rangle_{\ssr},
\qquad
\delta I_a(t)=I_a(t)-\langle I_a\rangle_{\ssr}.
\end{equation}
The output spectrum is the Fourier transform
\begin{equation}
S_{ab}^{\out}(\omega)
=
\int_{-\infty}^{\infty}
\dd\tau\, e^{\ii\omega\tau} C_{ab}(\tau).
\label{eq:app_spectrum_definition}
\end{equation}
The vacuum input part of the homodyne current gives the white shot-noise contribution.  For a single ideal homodyne current this contribution is $1$ in the real lock-in normalization used in the main text.

The system-dependent part is obtained from the quantum regression theorem.  For the homodyne current in Eq.~\eqref{eq:app_Xtheta_def}, define the current-insertion superoperator
\begin{equation}
\calB_\theta\rho
=
e^{-\ii\theta}L\rho
+
e^{\ii\theta}\rho L^\dagger .
\label{eq:app_Btheta_def}
\end{equation}
This superoperator represents the effect of inserting the earlier output-current operator into a two-time correlation function.  
To see this insertion rule, consider an infinitesimal output time bin
and write the homodyne increment as
\[
dY_\theta(t)=e^{-i\theta}dB_{\rm out}(t)
+e^{i\theta}dB_{\rm out}^\dagger(t).
\]
For a system state \(\rho\), the one-bin Stinespring map gives, to the
order needed here, field coherences
\[
V\rho V^\dagger
=
\cdots
+L\rho\,\sqrt{dt}\otimes |1\rangle\langle0|
+\rho L^\dagger\sqrt{dt}\otimes |0\rangle\langle1|
+\cdots .
\]
Using
\(dB_{\rm out}\simeq \sqrt{dt}\,|0\rangle\langle1|\) and
\(dB_{\rm out}^\dagger\simeq \sqrt{dt}\,|1\rangle\langle0|\),
and tracing over the output bin, one obtains
\[
\operatorname{Tr}_{\rm bin}\!\left[
(\mathbf 1\otimes dY_\theta)V\rho V^\dagger
\right]
=
\left(e^{-i\theta}L\rho+e^{i\theta}\rho L^\dagger\right)dt
=
B_\theta\rho\,dt .
\]
Thus insertion of the earlier homodyne increment prepares the
unnormalized system operator \(B_\theta\rho\,dt\). Propagation for a
time \(\tau>0\) is then given by the unperturbed semigroup
\(e^{\mathcal L\tau}\), and the later homodyne mean is
\(\operatorname{Tr}[X_\theta(\cdot)]dt\). Therefore
\[
\langle dY_\theta(t+\tau)dY_\theta(t)\rangle_{\rm ss}^{\rm sys}
=
\operatorname{Tr}\!\left[
X_\theta e^{\mathcal L\tau}B_\theta\rho_{\rm ss}
\right]dt^2 .
\]
Dividing by \(dt^2\) implies, for $\tau>0$,
\begin{equation}
\langle I_\theta(t+\tau) I_\theta(t)\rangle_{\ssr}^{\rm sys}
=
\Tr\!\left[
X_\theta e^{\calL\tau}
\calB_\theta\rho_{\ssr}
\right],
\label{eq:app_qrt_raw}
\end{equation}
where the superscript ``sys'' indicates the part beyond the instantaneous vacuum noise.  The disconnected contribution is removed by replacing $\calB_\theta\rho_{\ssr}$ by its projected version
\begin{equation}
\calQ(\calB_\theta\rho_{\ssr})
=
\calB_\theta\rho_{\ssr}
-
\rho_{\ssr}\Tr[\calB_\theta\rho_{\ssr}].
\end{equation}
Since
\begin{equation}
\Tr[\calB_\theta\rho_{\ssr}]
=
\Tr[X_\theta\rho_{\ssr}],
\end{equation}
this subtraction is precisely the subtraction of
$\langle I_\theta\rangle_{\ssr}^2$.

For $\tau>0$ the connected system part is therefore
\begin{equation}
C_\theta^{\rm sys}(\tau)
=
\Tr\!\left[
X_\theta e^{\calL\tau}
\calQ(\calB_\theta\rho_{\ssr})
\right].
\label{eq:app_positive_time_corr}
\end{equation}
The one-sided Fourier-Laplace transform gives the reduced resolvent:
\begin{align}
\int_0^\infty \dd\tau\,
e^{\ii\omega\tau}
C_\theta^{\rm sys}(\tau)
&=
\Tr\!\left[
X_\theta
\int_0^\infty \dd\tau\,
e^{(\calL+\ii\omega)\tau}
\calQ(\calB_\theta\rho_{\ssr})
\right]
\nonumber\\
&=
\Tr\!\left[
X_\theta
(-\ii\omega-\calL)^{-1}
\calQ(\calB_\theta\rho_{\ssr})
\right].
\label{eq:app_one_sided_resolvent}
\end{align}
For a real homodyne current, the negative-time part of the correlation function gives the complex conjugate of Eq.~\eqref{eq:app_one_sided_resolvent}.  Hence the full two-sided spectrum is
\begin{equation}
S_\theta^{\out}(\omega)
=
1+
2\Ree\,
\Tr\!\left[
X_\theta
(-\ii\omega-\calL)^{-1}
\calQ(\calB_\theta\rho_{\ssr})
\right].
\label{eq:app_homodyne_spectrum}
\end{equation}
This is the formula used in the homodyne examples.

For multiple real output currents, the same argument gives a matrix spectrum.  If
\begin{equation}
I_a(t)=\xi_a(t)+X_a(t),
\end{equation}
and if $\calB_a$ denotes the corresponding current-insertion superoperator, then
\begin{align}
S_{ab}^{\out}(\omega)
&=
D_{ab}
+
\Tr\!\left[
X_a(-\ii\omega-\calL)^{-1}
\calQ(\calB_b\rho_{\ssr})
\right]
\nonumber\\
&\quad
+
\Tr\!\left[
X_b(\ii\omega-\calL)^{-1}
\calQ(\calB_a\rho_{\ssr})
\right],
\label{eq:app_matrix_spectrum}
\end{align}
where $D_{ab}$ is the white-noise covariance of the input vacuum noises $\xi_a,\xi_b$.  For a single ideal homodyne current, $D_{\theta\theta}=1$, and Eq.~\eqref{eq:app_matrix_spectrum} reduces to Eq.~\eqref{eq:app_homodyne_spectrum}.

Specializing the multiple-current formula \eqref{eq:app_matrix_spectrum} to the two independent vacuum outputs of the truncated cat resonator gives \(D_{\alpha\beta}=\delta_{\alpha\beta}\) and
\begin{align*}
C_{\alpha\beta,N}^{\out}(\omega)
&=
\delta_{\alpha\beta}
+
\Tr\!\left[
X_{\alpha,N}(-\ii\omega-\calL_N)^{-1}
\calQ_N(\calB_{\beta,N}\rho_{\ssr,N})
\right]
\nonumber\\
&\quad+
\Tr\!\left[
X_{\beta,N}(\ii\omega-\calL_N)^{-1}
\calQ_N(\calB_{\alpha,N}\rho_{\ssr,N})
\right].
\end{align*}
This reproduces Eq.~\eqref{eq:kpo_complex_spectrum_matrix}, where \(X_{\alpha,N}\) and \(\calB_{\alpha,N}\) are defined in Eq.~\eqref{eq:kpo_Xalpha}.  Applying the block realification \(\mathsf B\) to every output-output and output-signal entry yields Eq.~\eqref{eq:kpo_real_matrices}.  This retains the cross-spectrum between the two homodyne records and is essential for the off-diagonal information block shown in Eq.~\eqref{eq:kpo_numeric_matrix}.

\bibliographystyle{iopart-num}
\bibliography{ref.bib}

\end{document}